\documentclass[sigconf]{acmart}

\AtBeginDocument{%
  \providecommand\BibTeX{{%
    \normalfont B\kern-0.5em{\scshape i\kern-0.25em b}\kern-0.8em\TeX}}}

\copyrightyear{2024}
\acmYear{2024}
\setcopyright{acmlicensed}\acmConference[IUI '24]{29th International Conference on Intelligent User Interfaces}{March 18--21, 2024}{Greenville, SC, USA}
\acmBooktitle{29th International Conference on Intelligent User Interfaces (IUI '24), March 18--21, 2024, Greenville, SC, USA}
\acmDOI{10.1145/3640543.3645159}
\acmISBN{979-8-4007-0508-3/24/03}
\usepackage{enumitem}
\usepackage{booktabs}
\usepackage{graphicx}
\usepackage{caption}
\usepackage{tabularx}
\usepackage{subcaption}
\usepackage{multirow}
\usepackage{array,ragged2e}
\newcolumntype{P}[1]{>{\RaggedRight\arraybackslash}p{#1}}

\usepackage{booktabs}% http://ctan.org/pkg/booktabs
\newcommand{\tabitem}{~~\llap{\textbullet}~~}

\begin{document}

% \title[ReviewFlow]{ReviewFlow: Intelligent Scaffolding to Support Academic Peer Reviewing}
\title{ReviewFlow: Intelligent Scaffolding to Support Academic Peer Reviewing}


\author{Lu Sun}
\affiliation{%
 \institution{University of California San Diego}
 \city{La Jolla}
 \state{CA}
 \country{USA}
 }
\author{Aaron Chan}
\affiliation{%
 \institution{University of California San Diego}
 \city{La Jolla}
 \state{CA}
 \country{USA}
 }
\author{Yun Seo Chang}
\affiliation{%
 \institution{University of California San Diego}
 \city{La Jolla}
 \state{CA}
 \country{USA}
 }
\author{Steven P. Dow}
\affiliation{%
 \institution{University of California San Diego}
 \city{La Jolla}
 \state{CA}
 \country{USA}
 }

% \renewcommand{\shortauthors}{}

\begin{abstract}

Peer review is a cornerstone of science. Research communities conduct peer reviews to assess contributions and to improve the overall quality of science work. Every year, new community members are recruited as peer reviewers for the first time. 
How could technology help novices adhere to their community's practices and standards for peer reviewing? To better understand peer review practices and challenges, we conducted a formative study with 10 novices and 10 experts. We found that many experts adopt a workflow of annotating, note-taking, and synthesizing notes into well-justified reviews that align with community standards. Novices lack timely guidance on how to read and assess submissions and how to structure paper reviews. To support the peer review process, we developed ReviewFlow -- an AI-driven workflow that scaffolds novices with contextual reflections to critique and annotate submissions, in-situ knowledge support to assess novelty, and notes-to-outline synthesis to help align peer reviews with community expectations.
In a within-subjects experiment, 16 inexperienced reviewers wrote reviews in two conditions: using ReviewFlow and using a baseline environment with minimal guidance. With ReviewFlow, participants produced more comprehensive reviews, identifying more pros and cons. However, they still struggled to provide actionable suggestions to address the weaknesses. While participants appreciated the streamlined process support from ReviewFlow, they also expressed concerns about using AI as part of the scientific review process. We discuss the implications of using AI to scaffold the peer review process on scientific work and beyond.
\end{abstract}

\begin{CCSXML}
<ccs2012>
   <concept>
       <concept_id>10003120.10003121.10011748</concept_id>
       <concept_desc>Human-centered computing~Empirical studies in HCI</concept_desc>
       <concept_significance>500</concept_significance>
       </concept>
 </ccs2012>
\end{CCSXML}

\ccsdesc[500]{Human-centered computing~Empirical studies in HCI}

\keywords{intelligent scaffolding, academic peer review, Large Language Models LLMs)} %visualization} %news reading,


\maketitle
\section{Introduction}

% peer review is important, but more submission increase burden of reviews -> more novice reviewers are need to be involved
Peer review is a cornerstone of academic research, ensuring the quality, credibility, and reliability of scientific research~\cite{price2017computational}. 
The peer review process seeks to assess whether submissions contribute new knowledge to a research community and generate feedback that helps authors improve the quality of their work ~\cite{jefferson2002effects,shah2022overview}. Many communities are seeing a rapid increase in submissions~\cite{mccook2006peer,stelmakh2019peerreview4all,shah2022overview,arous2021peer,stelmakh2021novice}; while this could be seen as an indicator of scientific progress, it also has increased the pressure on reviewers. To meet increased demand, many research communities recruit a significant number of first-time reviewers or ACs (Associate Chairs) for each review cycle. For example, in 2023, the ACM CHI Conference on Human Factors in Computing Systems reported that more than 50\% of ACs were first-time ACs~\cite{chi23report}.

% Faced with the reviewers turn-over, novices may have significant changes, we argue that scaffolding can help
Peer reviewing is a complicated and challenging task. Reviewers need to understand the paper, evaluate the scientific content using domain knowledge, make a fair decision, and compose a comprehensive review to communicate their assessments and recommendations~\cite{shah2022overview}. It is a task that requires critical thinking, a deep understanding of the subject, and the ability to provide constructive feedback. Conferences and journals need to ensure that first-time reviewers meet the standards of the research community. 

\begin{figure*}[t]
    \centering
    \includegraphics[width=\textwidth]{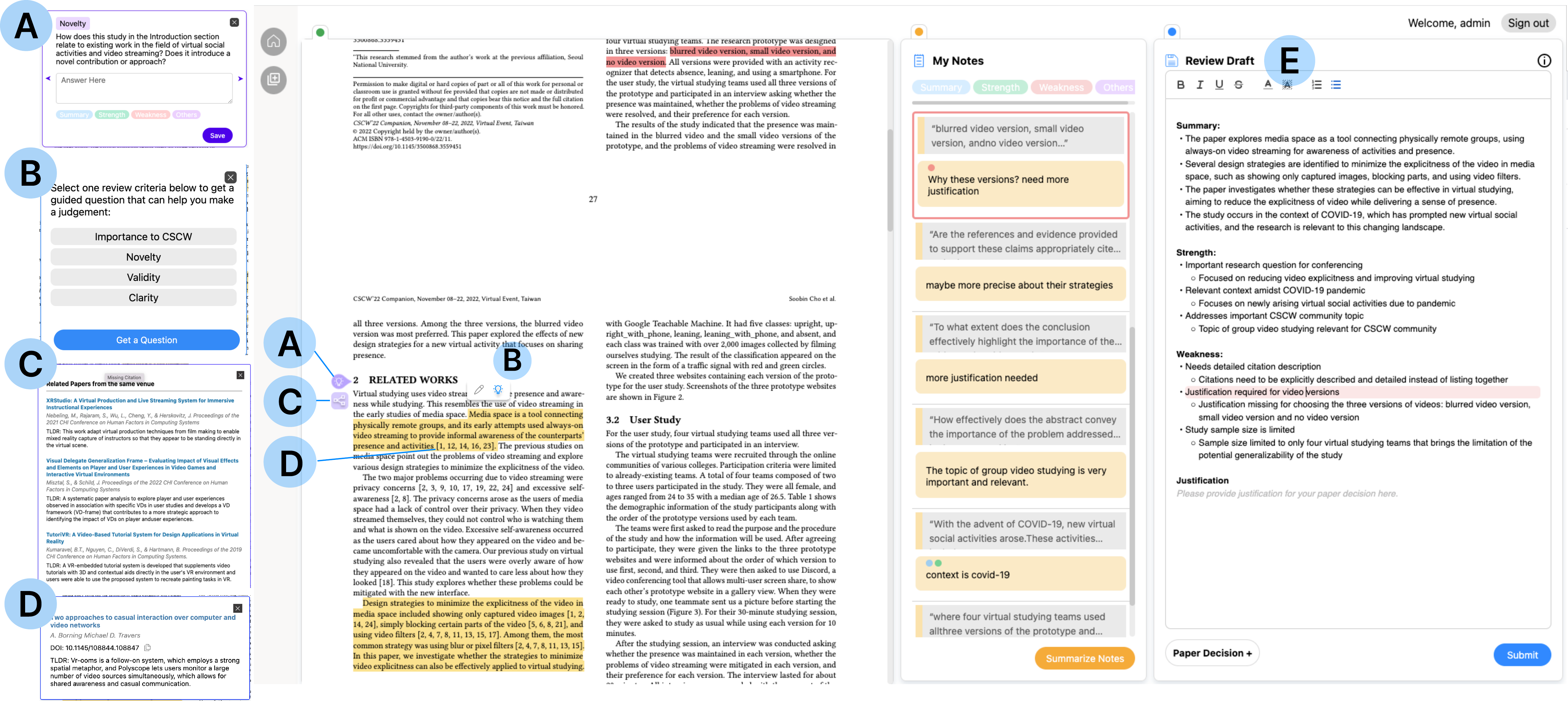}
    \caption{ReviewFlow interface on an example paper. Users can (A) receive section-level contextual cues guided by community criteria; (B) request phrase-level contextual cues adapted to highlight paper content; (C) click the citation to get an in-situ summarization; (D) check the recommended citations not currently cited by the paper; and (E) click to summarize the notes into a high-level outline or expand into a detailed outline }
    \label{fig:system_overview}
\end{figure*}

% scaffoldings have been used in different contexts, reading and writing. Here we use AI scaffolding that can be contextualized and integrated into the review workflow. 
% Motify illustrates that scaffolds extracted from expert examples help novices create video stories~\cite{kim2015motif}.
One approach to preparing novices for any complex task is to provide scaffolding, a strategy used in educational settings to support learning and mastery~\cite{reiser2004scaffolding,collins2006cognitive}. 
Scaffolding takes many forms, including examples of prior work, templates, or hints to help novices think and potentially perform on par with experts~\cite{holton2006scaffolding,saye2002scaffolding}.
Previous studies found that scaffolded examples and templates can even help learners perform similarly to experts in terms of feedback quality~\cite{yuan2016almost}. 
For instance, the LetterSmith project developed an approach to aid writing called ``scaffolded annotation'' which provided key components of the writing tasks and annotated expert examples, and helped professional writing students improve the quality of their early-stage drafts ~\cite{hui2023lettersmith}. 

Advances in AI and large-language models (LLMs) in particular, have the potential to make scaffolding even more effective because they can detect and adapt to the user's work context~\cite{rachatasumrit2022citeread}. 
For example, the CReBot project leverages LLMs to generate and place questions within an academic paper to help induce critical thinking while reading~\cite{peng2022crebot}; results showed that CReBot helped novices read and comprehend the paper content. 
While LLMs show great potential, they are also known to hallucinate~\cite{bang2023multitask} and perpetuate systemic biases~\cite{huang2019reducing,jakesch2023co} that could negatively impact the effectiveness of scaffolding.
While intelligent scaffolding has proven valuable for well-scoped tasks, academic peer review involves reading submissions with a critical eye, synthesizing knowledge, and making a well-justified judgment for acceptance or rejection. Our research investigates how intelligent scaffolds can guide a complex, multi-faceted workflow for academic peer reviewing without biasing the ultimate decisions.
This paper explores two key research questions:  What are the challenges faced by novices and the strategies adopted by experts during the peer review process? How can we intelligently scaffold the peer review process?

To explore how we might use AI scaffolding to support the peer review process, we first conducted a formative study to understand what novices see as key challenges and how experts approach this task. We interviewed 10 novices with limited prior experience writing peer reviews to articulate their obstacles during the process. They expressed challenges around the lack of sufficient guidance on how to write a well-structured peer review peer review and how to make judgments about the paper's quality. Furthermore, we conducted observational studies with 10 experts to ask them to write a peer review for a short paper and to think aloud as they complete their workflow. Then, we invited experts to provide their perspectives on using AI to support the tasks and express their concerns. We found that many experts adopt a workflow involving critical reading, annotating, note-taking, and synthesizing a well-justified review that conforms to community guidelines.

% should we highlight each features novelty?
Based on these insights, we developed a prototype -- ``ReviewFlow''\footnote{ReviewFlow Code Repository: https://github.com/LusunHCI/ReviewFlow.git}, a platform for writing peer reviews that incorporates intelligent scaffolding to support a workflow for inexperienced reviewers. ReviewFlow incorporates a range of features to facilitate the review process: (1) \textbf{Contextual cues} are embedded questions that help reviewers reflect on the paper. ReviewFlow includes section-level cues guided by the community's review criteria as well as phrase-level cues that adapt to the paper's content. (2) \textbf{In-situ citation recommendations} show relevant but non-cited papers that may help reviews assess the novelty compared to existing work. (3) \textbf{Notes-to-outline synthesis} guides reviewers to organize notes and structure reviews to align with community standards. ReviewFlow gathers all the notes left by the reviewer and leverages an LLM to summarize notes into a high-level outline. Reviewers can revise and add detail to the outline while adhering to community standards.

% Conducted within subject experiment,  we measured xxx and found xxx. participants also revealed concerns xxx
We conducted a within-subjects study to evaluate ReviewFlow where (N=16) participants --- with little to no experience as reviewers --- wrote reviews for two short papers in a counterbalanced manner: one using the ReviewFlow with all scaffolding features and one using the baseline interface with only traditional forms of guidance (e.g., review rubrics and an example review).  We found that novice reviewers wrote significantly more structured and more comprehensive reviews in the ReviewFlow system than in the Baseline system, as evaluated by experts. Novice reviewers wrote slightly more constructive reviews in the ReviewFlow system, but the difference was not significant. Reviewers called out more weaknesses in the paper using ReviewFlow, but they still struggled to provide actionable suggestions for the authors to address the weaknesses.

% Contribution:
Our paper offers several contributions:
First, a formative study revealed that novices lack opportune guidance on key considerations and expectations and uncovered common practices adopted by experts in the review process.
Second, we developed ReviewFlow to model experts' workflow for peer reviewing while also leveraging LLMs to provide contextual cues, in-situ knowledge recommendations, and notes-to-outline synthesis.
Third, we gained empirical insights from a within-subjects study with 16 participants which revealed how intelligent scaffolding can help novices write well-structured and comprehensive reviews.
\section{Related Work}

\subsection{Practices and challenges related to academic peer reviewing}
As an important step for ensuring the scientific quality of work, peer review has been adopted by most journals and conferences~\cite{price2017computational}. In a typical conference review process, each reviewer needs to evaluate the paper's quality, make an acceptance decision and provide reviews for their assigned papers~\cite{shah2018design,shah2022overview}. The reviewers are usually experts in the area who have fruitful experience and knowledge to assess or evaluate the quality and contribution of the paper. After reviewers submit the review, a discussion takes place between reviewers and a meta-reviewer, who will carry out the final decision on the acceptance of the paper. 

The number of papers in research communities has increased exponentially in recent years~\cite{wang_reviewrobot_2020}. While this may be positively viewed as an acceleration of scientific progress, the disparity between growth rates of the submission and reviewer pools also creates more burden for reviewers~\cite {mccook2006peer,stelmakh2019peerreview4all,stelmakh2021novice,shah2022overview}. To avoid overloading reviewers, conferences need to find new sources of reviewers as there are not enough experienced reviewers to review all papers~\cite{stelmakh2021novice}. These novice and junior reviewers constitute a large fraction of the reviewer pool in computer science conferences~\cite{stelmakh2021prior}. 
For example, in 2023, the ACM CHI Conference on Human Factors in Computing Systems reported that more than 50\% of ACs who reviewed papers are first-time ACs~\cite{chi23report}. Given this large fraction, conferences need to ensure that newly added junior reviewers do not compromise the quality of the process, that is, are able to write reviews of quality comparable to the experienced reviewers. 
 % For example, using data on the structure of the reviewer pool of the ICML 2020 conference, researchers estimated that approximately 35\% of the reviewers were novice reviewers who are junior individuals who self-nominated and satisfied the screening requirements of having one or two papers published in some top ML venues~\cite{stelmakh2021prior}. 

A previous study explored and compared the reviews written by experienced reviewers versus junior reviewers and the study showed that junior reviewers were slightly harsher in scoring the clarity of the submissions~\cite{shah2018design,stelmakh2021novice}. In the meantime, other works provide empirical evidence that junior reviewers are more critical than their senior counterparts and reveal that graduate students' review comments are not very useful~\cite{mogul2013towards,patat2019distributed}. Faced with these doubts and challenges, helping novice junior reviewers to write a high-quality review becomes crucial.

Reviewing is a time-consuming and mentally demanding task~\cite{wang_reviewrobot_2020,shah2022overview}. A constructive and comprehensive review can improve the quality of the paper, while a bad, random, dismissive, or biased review brings frustration and anger to authors~\cite{wang_reviewrobot_2020}. To provide a high-quality review, reviewers need to go through a multi-step workflow, including understanding the contribution of the paper, accessing the merit of scientific contribution and providing an evaluation together with a comprehensive written review~\cite{smith2006peer,jefferson2002effects,xiong_automatically_2011,langford2015arbitrariness,hua2019argument,cheng2020ape,gao2019does,wang_reviewrobot_2020,yuan2021can}. 
To make a fair judgment, reviewers need to have enough background knowledge to grasp the main idea of the paper and evaluate its contribution. More importantly, reviewers need to equip critical thinking skills to think deeply about the author's judgments, like whether the claims are reasonable and why the approaches are chosen~\cite{facione1990critical,moore2013critical}. A typical peer review not only contains the paper summary and its contribution but also raises weaknesses from different aspects together with constructive feedback or thought-provoking questions. During this process, it requires readers to actively analyze, synthesize, and evaluate the paper content~\cite{moore2013critical}. 

Researchers are typically trained extensively in conducting the research itself, but they often lack formal instructions in the peer review process. Hence, it becomes even more challenging for inexperienced junior reviewers to gain expertise quickly~\cite{shah2022overview}. Existing research explored instructional methods to ``teach'' or ``train'' junior reviewers. A previous study provided a training video and found that it increased the inter-reviewer agreement, alignment with the scoring rubric, and the amount of time reading the review criteria~\cite{sattler2015grant}. Another study offers novice reviewers a more guided introduction to the different stages of the reviewing process, such as how to lead a discussion among reviewers, to help novices write better reviews. The results showed that with this guidance on the reviewing stages, novice reviewers could deliver more ``above expectation'' reviews~\cite{stelmakh2021novice}. However, their guidance is only limited to introducing the different parts of the reviewing process, such as rebuttal and discussion, and providing novices opportunities to ask expert questions on the general process~\cite{stelmakh2021novice}. Outside of the general review process, academic peer review is a complex activity that involves both understanding a paper submission's stated contributions and evaluating whether the paper crosses an acceptable threshold for the research community.  To explore how we might support the peer review process, we start with a formative study to understand how experts approach this task and what novices see as key challenges.  

Previous research has used computational methods to provide support to streamline several parts of the peer review process, such as matching submissions with appropriate reviewers or assessing review quality~\cite{bartoli2016your,wang_reviewrobot_2020,hua2019argument,yuan2021can,arous2021peer,shah2018design}. However, fewer empirical studies that attempt to scaffold the entire workflow for reviewing academic papers, which includes reading, note-taking, evaluating, decision making, and synthesizing this into a written review. 
% Moreover, past studies find that there are no easily identifiable types of formal training or experience that could predict reviewers’
% review quality~\cite{callaham2007relationship}

\subsection{Scaffolding strategies for complex cognitive tasks}
To help novices improve problem-solving skills in complex cognitive tasks, cognitive apprenticeship introduces several strategies, including modeling, coaching, scaffolding, and reflection~\cite{vygotsky1978mind,collins2006cognitive}. Scaffolding is instructional support provided by experts to promote learning, especially when concepts and skills are being first introduced to novice students~\cite {reiser2004scaffolding,collins2006cognitive}. These supports include advanced organizers, modeling, worked examples, concept maps, explanations, handouts, and prompts~\cite{oakes2018instructional,nesbit2013concept,atkinson2000learning,charney1995learning,bui2015enhancing,collins2006cognitive}. 
Previous research shows that effective scaffolding can help novices perform work nearly as well as experts~\cite{yuan2016almost,hui2023lettersmith, hui2018introassist}. When scaffolding is mediated by technology, including AI-based methods, it creates more opportunities for instructors and learners but also brings more challenges in making the scaffolding contextualized, adaptive, and effective~\cite{hannafin1999open}.

Researchers used prompts and guided questions to scaffold learners in the paper reading process~\cite{peng2022crebot,yuan2023critrainer,chen2022marvista}. To facilitate critical paper reading, researchers developed CReBot which interactively asks section-level critical thinking questions for routine paper readers and novice readers. Results showed that the interactive question prompts CReBot provided might not be better than static guidelines for beginners to conduct critical thinking. Furthermore, researchers developed CriTrainer which can adaptively provide questions in the reading process together with hints and feedback to help readers critically think and comprehend the paper content. Interestingly, on the opposite of CReBot, their result showed that CriTrainer can improve learners' ability to raise understandable, relevant, and critical questions after the training sessions. Their results highlighted the benefits of its text-specific critical thinking questions provided by the system. However, guided questions used in CReBot and CriTrainer are both template-based and did not fully use the user-selected content on the paper.

Existing research also used existing examples to scaffold complex writing processes~\cite{yuan2016almost,hui2023lettersmith, hui2018introassist}. Scholars found that scaffolding can help students learn about form and organization by analyzing the examples and templates~\cite{doan2021teaching,collins2006cognitive}. In the context of writing introductory help requests, providing high-quality examples and expert-informed templates can increase learning and writing quality~\cite{hui2018introassist}. Another writing support system used ``scaffolded annotation'' that broke down examples into each component to help professional writing students improve their early-stage drafts~\cite{hui2023lettersmith}. In traditional instruction scenarios, experts take a large amount of time to curate examples, create guidance, or author rubrics~\cite{reiser2004scaffolding}. However, experts who created examples still faced the challenges of effectively adapting and contextualizing into the current learning step.

\subsection{Leveraging AI scaffolding to support writing}

Advances in AI and LLMs provide opportunities to provide efficient and context-specific scaffolding in the learning process~\cite{ippolito2022creative,chung2022talebrush,clark2018creative,gero2019metaphoria,gero2019stylistic,rahman_mixtape_2020,lee_coauthor_2022,gero_sparks_2021,ippolito2022creative,mirowski_co-writing_2022}. For example, TaleBrush allowed users to create a story with AI through sketching to aid the planning of writing. Then it enabled writers to generate diverse storylines and interactively refine them~\cite{chung2022talebrush}. Another system Wordcraft explored how to support users collaborating with generative language models to co-write a story~\cite{yuan2022wordcraft}. Spark used a language model to generate prompts related to a scientific concept to facilitate scientific writing~\cite{gero_sparks_2021}. However, as far as we know, none of the studies that used scaffolding strategies focused on the context of conference peer review writing. Our research explores how we might guide a complex, multi-faceted workflow like peer reviewing and the potential role of AI in creating contextually adaptive scaffolds.

Writing is a complex, iterative process~\cite{flower1981cognitive}. The Hayes model describes the cognitive processes an individual writer engages in during the process of writing~\cite{flower1981cognitive,hayes2012modeling,gero2022design}. In the cognitive process of writing, there are three major components: planning, translating, and reviewing, as shown in Figure~\ref{fig:feature-writing}. Several systems are developed to facilitate different stages of writing~\cite{bhat2023interacting,zhang2023visar}. For example, VISAR is an AI-enabled writing assistant that helps writers brainstorm and revise hierarchical goals and organize argument structures in the planning stage. To facilitate the iterative planning and revising process in writing, intelligent systems further use the chain of thoughts (COT) prompting method to break down the large problem into step-by-step prompts~\cite{wei2022chain}. Specifically, the Re3 framework and DOC framework used the COT approach to decompose the writing tasks where they first generate an outline and then automatically turn the outline into the story generation~\cite{yang2022re3,yang2022doc}. Their evaluation demonstrated that the decomposition approach can improve the coherence of long story generation and is highly controllable where humans can control the story generation by modifying the outlines. 

Control and agency are extremely important in the conference peer review writing process, as human reviewers should play the role of driving the writing process. While the development of LLMs can provide scaffolding opportunities for this writing process, it is important to address the concerns and limitations of LLMs in this context~\cite{fok2023can}.  
The prior work that aligns most closely with the concept of applying AI techniques to reviewing academic papers that a machine model that automatically generates feedback using LLMs. Results showed that LLM feedback could benefit researchers in earlier stages of manuscript preparation while researchers struggle with an in-depth critique of study methods~\cite{liang2023large}. Instead of using an automatic method, humans should be in the loop to drive and control the writing process~\cite{fok2023can}.
Drawing on these insights, we designed the ReviewFlow system not to automate any parts of the process, but rather to scaffold key considerations and to give novices agency over how to apply machine-generated language suggestions.  

% \vspace{-1em}
% \begin{figure}
%     \centering
%     \includegraphics[width=0.5\textwidth]{figures/writing-process-cognitive.png}
%     \caption{The cognitive process model for writing, as proposed by Flower and Hayes.}
%     \label{fig:writing-process}
% \end{figure}

\section{Formative Studies}
Before we developed our system to support academic peer reviewing, we conducted two formative studies.  
First, we interviewed ten novice reviewers to understand how they approached this task for the first time and their perceived challenges (represented as C). Second, we conducted observational studies with ten experienced reviewers where we invited them to write a peer review on a selected paper to capture their common practices and workflows. Last, based on the findings, we proposed design goals (represented as DG) for supporting novice reviewers.

\subsection{Methods}
\subsubsection{Novice Interview Study} \hfill \break
We conducted semi-structured interviews with 10 novice reviewers (N1-N10) with relatively little experience with writing academic peer reviews (ranging from only reviewing once before and having less than 2 years of experience). Participants were recruited through mailing lists and social media posts. The participants (4 female and 6 male, average age of 25.5 years) came from diverse research fields including Human-Computer Interaction, AI, Cognitive Science, and Computer Security. Interviews were conducted remotely by the lead author and lasted around 30 minutes.

In the interview, we first asked open-ended questions about their perceived obstacles and challenges of conducting academic peer review. We then provided scenarios of potential features for peer review scaffolding to elicit reactions from novice reviewers. These scenarios described common potential situations faced by first-time reviewers and were designed to prompt the participants to share their needs in a real-world situation~\cite{gaver1999design}. For example, ``Mary is a first-year graduate student doing Human-Computer Interaction (HCI) research. While she has submitted a couple of papers before she received reviews, this will be her first time as a reviewer. She struggled to make the review constructive for the authors''.  Interviews were recorded with participants’ permission and were transcribed. Two researchers from the team went through the transcripts and coded themes using thematic analysis~\cite{braun2006using}. Through multiple iterations along with periodic discussions, the coding led to the major themes of challenges below.

\subsubsection{Experts Observational Study} \hfill \break
We conducted observational studies with 10 experts reviewers(E1-E10) with at least 5 years of experience in writing academic peer reviews, to learn about their best practices~\cite{karen2017contextual}. Participants were recruited through in-person invitations, email lists, and social media posts. Participants (4 female and 6 male, average age of 30.1 years) came from different fields of computer science, including HCI, AI, programming languages, learning science, computer security, and accessibility. Observational studies were conducted remotely by the lead author and lasted around 90 minutes.

During the observational study, we first spent 20 minutes asking about their experiences and challenges with peer reviewing. Then, the team asked the participant to write a peer review in a Google document for one of 5 different short papers (less than 4 pages) in 60 minutes. To find suitable short papers, the team first collected the participants' research interest descriptions from their websites and used these keywords to search on Semantic Scholar~\cite{semanticscholar}. We filtered out the papers longer than 5 pages, ranked them by relevance, and selected the top 5 as the participant's options for the study. 

In the end, we provided a series of design probes for the novices to capture their reactions. Drawing insights from previous literature around the intelligent support on reading and writing, three researchers on our team took multiple rounds of iteration and group discussions to develop the six design probes~\cite{zhang2023visar,lo2023semantic}. These design problems are visualized in Figma. The design probes included: (1) scaffold annotation with community curated tags and filters, (2) reflection questions (expert-authored generic questions versus AI-generated specific reflection questions on each section or highlighted sentences), (3) extractive summarization and generated explanations to facilitate paper reading, (4) in-situ citation recommendation where the system provides a summary from the cited paper and recommend potential missing citations, (5) review draft generation, (6) mapping back the source of review draft from the paper content and visualize the location of the source to help reviewers revise their draft. The interviews were recorded and then transcribed using a machine transcription service. The research team took the same analysis procedure as the novice study. 

% Last, we provided the same design probes to the experts to elicit what scaffolding strategies could benefit the novices the most. 

\subsection{Findings}

\subsubsection{Novice reviewers felt they lacked guidance in evaluating the paper and writing structured and constructive reviews}

% C1 Lacks opportune considerations for assessment
% C2 Lacks domain knowledge for evaluating novelty
% C3 Lacks guidance on process and structure
% C4 Lacks effective models for tone and expectation
The novice participants reported that their prior attempts at reviewing papers were cognitively demanding and took an average of 6.4 hours.
When asked about challenges, 6/10 mentioned that they \textit{struggled to fully understand the background knowledge and existing work and did not feel confident about assessing the paper's novelty (C2)}. 4/10 mentioned that they \textit{lack opportune considerations for assessment (C1)}. Specifically, one participant mentioned that they would love to have ``co-reviewers who have already read the paper and known the specifics to guide me through the evaluation process''(N5). In addition, 4/10 highlighted that they \textit{need more guidance on how to structure issues during the writing process(C3)}. 3/10 explicitly mentioned that they want to receive some feedback from experts, other reviewers, or authors, to \textit{make sure the review is in the right tone and meets the expectation (C4)}. Specifically, one participant mentioned ``I am not sure whether I covered all the necessary points or whether authors will perceive the reviews as useful''[N7]. Another participant is worried about being ``impolite or harsh''[N9].

We presented novices with the design scenarios one by one and asked them to rate the degree to which they resonate with each scenario (1 = Not resonate at all to 5 = Strongly resonate with the situation). 
The top two situations that resonated with novices the most were: ``the novice reviewer struggles to write a high-quality review'' (3.8/5) and ``the novice reviewer spends a lot of time reading a paper but doesn't know how to evaluate the paper'' (3.8/5). 
% In addition, participants also resonate with the situation that ``the novice reviewer takes several hours to conduct extensive background research to understand the paper''(3.6/5). 

\begin{figure*}[htb]
    \centering
    \includegraphics[width=\textwidth]{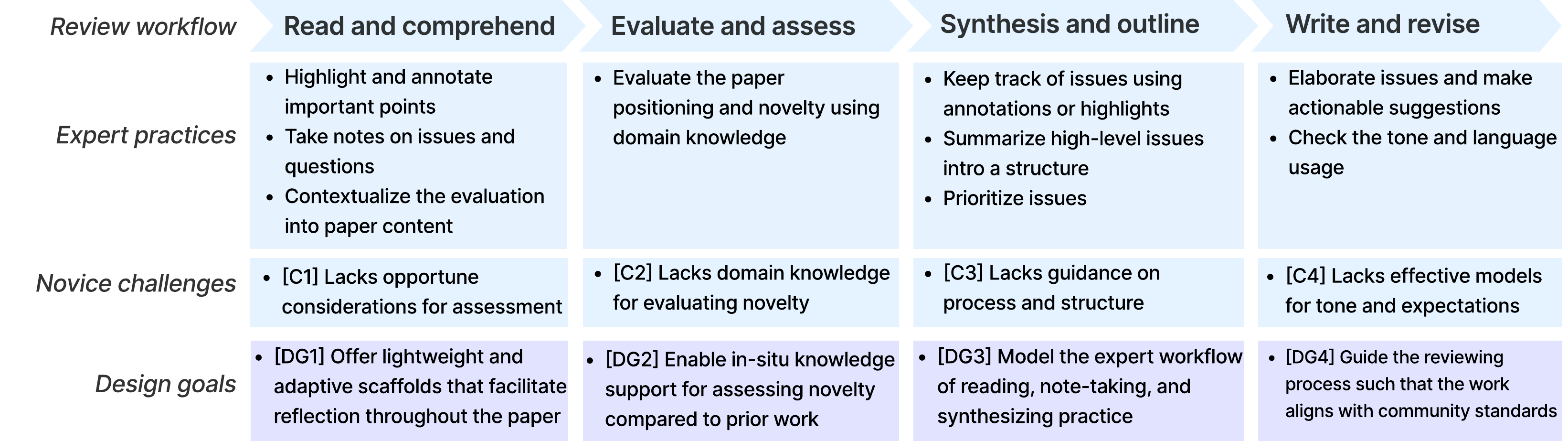}
    \caption{Experts' workflow in academic peer review, along with experts' practices, novices challenges, and design goals for each stage}
    \label{fig:workflow}
\end{figure*}

\subsubsection{Experienced reviewers adopt a workflow of sense-making, annotating, and synthesizing notes}

Experienced reviewers reported that they spent 4.75 hours to review one paper. In the observational study with 10 experts, we synthesized the common workflow adapted by experts in the reviewing process. As an initial step, reviewers read and comprehend the paper's content. While reading through the paper, all experts (10/10) highlighted some content, annotated sentences or paragraphs, and took some notes. The format and style of notes varied between different reviewers. We observed some reviewers using symbols while others used phrases or short sentences. After reading and annotating the content, reviewers start to re-read the notes to evaluate the paper's quality. 3 out of 10 experts went back to check the introduction and related work to assess the novelty of the paper. 

After finishing reading the paper, instead of directly editing the review draft, experts created high-level headers or topics,  such as ``lack clarity on study design'', that summarized the paper's weaknesses and strengths as well as prioritized the concerns (7/10). For instance, one expert reflected on the review process as ``I will make a lot of annotations and then I will just review them section by section. At the same time, thinking about what is the biggest concern in terms of the overall research novelty''(E6). We observed that 7 participants wrote the review in the following structure: 
summary of the paper, strengths or contribution of the paper, two to three weaknesses of the paper, and end the review with decision justification and general recommendations. Some experts will then list bullet points under each topic together with questions that they would like to ask the author. Last, experts compiled these comments and bullet points into a complete draft and revised them two to three times to offer suggestions and make it more constructive. Figure~\ref{fig:workflow} represents the expert's review workflow together with their practices.

\subsubsection{Experienced reviewers stressed the importance of specific and contextual guidance for scaffolding}

Among these features, 5 experts reflected that the contextual cues can be helpful, especially for novice reviewers. One participant preferred the AI-generated cues on the selected paper content and explained that ``I like to have some capability of freedom, but I think this will be super useful for novice reviewers who don't know how to review. But for people who have reviewed for so many years, those guidances for conferences are pretty much the same'' [E6]. We provided experts with two sets of tags to select. The first set of tags is designed based on the review structure that includes ``summary of the paper'', ``strength'', ``weakness'' and ``others''. The second set of tags is designed using community review criteria that include ``relevance'', ``novelty'', ``validity'', ``clarity''. Most experts preferred the first set of tags (7/10). 2 participants mentioned that they were concerned the experts' authored cues might be too similar to the existing guidelines. Hence, they suggested the contextual question can provide better guidance for novices.

9 out of 10 experts mentioned their preference for the in-situ citation support to provide summary and recommendations. They reflected that this in-situ knowledge support can ``raise awareness on unknown work'' [E5]. 9 out of 10 participants expressed their concerns about the summarization feature, as they don't trust the AI's ability to identify important information. Instead, they think reviewers should have control over the reading process. Specifically, one expert mentioned ``I think we are also reviewing the style of writing, or how something is communicated and logically connected between each paragraph or each sentence. So I think there is value with actually reading everything to get the message behind the paragraphs'' [E3]. Participants shared their opinions on using AI in the review process and all of them agreed that LLMs should not generate the review on the fly, and instead, human experts should drive the process since the limitation of LLMs can bias human experts and lead to over-reliance on the use of AI.

Based on the interview on the design probes, we summarize the following design considerations mentioned by novices and experts. The support should be lightweight and not distract from the current review flow. While involving AI in the review process, AI should not go too far to bias or lead the thinking process. Human reviewers should still preserve agency in reading, writing, and decision-making. Faced with the limitations of current LLMs, intelligent systems should try to avoid hallucinations and provide enough opportunities for fact-checking.

\subsection{Formative Study Discussion}
From the interview study with novices and the observational study with experts, we identified a range of challenges perceived by novices as well as insights on expert practices (see Figure~\ref{fig:workflow}).  Novices lack confidence in identifying the novelty of research based on prior research and in knowing how to structure a peer review that meets community standards. Experienced reviewers tend to avoid biasing the decision-making and preserve agency in reviewing.  The juxtaposition of these novice challenges and the expert practices suggests four core design goals for intelligent scaffolding:

\textbf{DG1: Offer lightweight and adaptive scaffolds that facilitate reflection throughout the paper}
Novices highlighted their need for guidance in critically evaluating papers from various perspectives, expressing uncertainty about the key points to focus on (C1). In educational settings, instructors typically serve as scaffolds when introducing new concepts or knowledge to novice learners~\cite{collins2006cognitive}. Prior studies in scientific paper reading developed methods to offer contextual hints or guided questions that encouraged reflection and critical thinking~\cite{august2022paper,rachatasumrit2022citeread}. For example, Paper Plain provided a collection of key questions that guide readers to answering passages and plain language summaries of those passages~\cite{august2022paper}. Similarly, in the context of our study, the objective is to provide guidance comparable to that of expert peer reviewers. The emphasis is on delivering locally relevant questions that help users think critically about each section of the paper.

\textbf{DG2: Enable in-situ knowledge support for assessing novelty compared to prior work}
Novices struggled with insufficient background knowledge for evaluating papers (C2). Providing knowledge support in situ can externalize the user's working memory, aid in sense-making, and facilitate a swift reviewing and resumption of task contexts~\cite{kuznetsov2022fuse}. Prior studies in scientific literature review highlight the importance of in-situ knowledge support~\cite{kang2022threddy,chang2023citesee}. To help novice reviewers who lack background knowledge to evaluate the novelty while reading the paper, our goal is to enable in-situ knowledge support by extracting the abstract of the cited paper.

\textbf{DG3: Model the expert workflow of reading, evaluating, and synthesizing practiced by experienced reviewers}
Novices expressed their need for more specific guidance from experts in terms of their process and structure (C3). 
Prior work showed that surfaced expert practices can better structure and scaffold the process for novices~\cite{kim2015motif}.
Motify illustrates that storytelling patterns extracted from expert stories can be used to effectively scaffold novices to create video stories~\cite{kim2015motif}. In the context of writing introductory help requests, providing high-quality examples and expert-informed templates can increase learning and writing quality~\cite{hui2018introassist}. Our approach involves the modeling of the expert workflow encompassing reading, evaluating, synthesizing, and revising. This model is then employed to structure the peer review process for novice reviewers.

\textbf{DG4: Guide the reviewing process such that the work aligns with community standards}
Novices reported that they lacked effective models for tone and expectation (C4). More guidance on the expectation of the review can help reviewers to reflect on their tone and structure.  In addition, experts raised valid concerns about the potential for intelligent tools to introduce biases or influence the thinking process. Existing studies have highlighted apprehensions regarding AI exhibiting biases in human-AI collaboration tasks~\cite{kawakami2022care,poursabzi2021manipulating}. Notably, research has shown that providing information, including explanations, generated by AI has the potential to mislead users in decision-making~\cite{lakkaraju2020fool,goyal2023else,buccinca2021trust}. Therefore, a key design objective is to steer clear of biasing the decision-making process and focus solely on encouraging justifications. This approach aims to assist novice reviewers in producing high-quality and original reviews without compromising the integrity of the evaluation process.
\section{ReviewFlow}

We developed ReviewFlow which employed AI-driven scaffolding strategies naturally into the review workflow to support novice reviewers to gain expertise in conference peer reviewing.

\subsection{Key features}
\subsubsection{\textbf{DG1: Contextual cues}}

\begin{figure*}[htb]
    \centering
    \includegraphics[width=\textwidth]{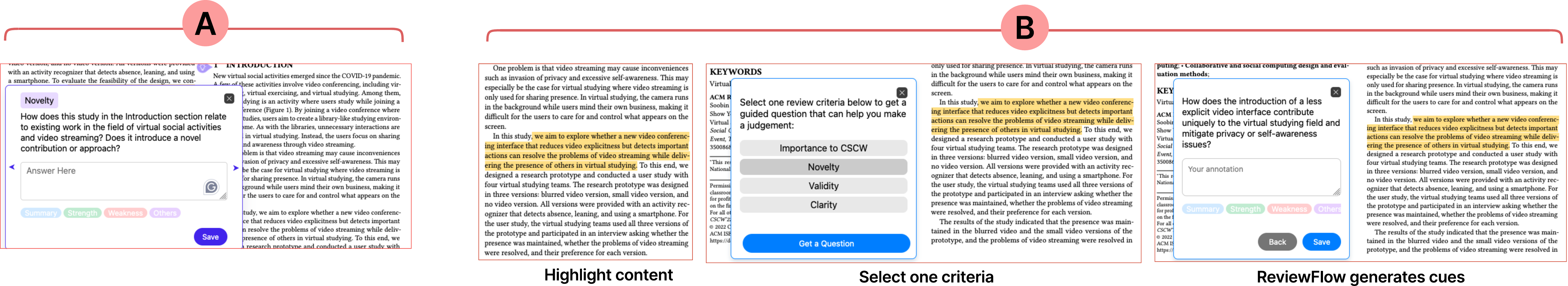}
    \caption{Contextual cues. ReviewFlow provides (A) section-level cues guided by community criteria and (B) phrase-level cues adapted to user highlighted content and selected criteria.}
    \label{fig:feature-reflection}
\end{figure*}

Researchers used prompts and guided questions to scaffold learners in the paper reading process~\cite{peng2022crebot,yuan2023critrainer,chen2022marvista}. Understanding the paper and critically reflecting on the content is essential for decision-making and review writing in the later stage. Prior research showed that providing questions can increase user engagement in actively searching for information. Prior systems mostly used template-based guided questions to engage readers~\cite{peng2022crebot}. To \textit{offer lightweight and adaptive scaffolds that facilitate reflection throughout the paper [DG1]}, ReviewFlow provides two types of contextual cues for novices, as shown in Figure~\ref{fig:feature-reflection}. One type is \textbf{section-level cues guided by community criteria} which are adapted based on each paper section's content together with the review criteria. For section-level reflection cues in Figure~\ref{fig:feature-reflection}-A), it takes the entire section, the paper abstract, and the community criteria into account to generate questions. Users can either read these questions before they read the section to obtain contextual guidance or after to reflect on the section's content. Another type is \textbf{phrase-level cues adapted to paper content}. As shown in Figure~\ref{fig:feature-reflection}-B, users have the freedom to highlight the content that they would like to reflect on at the phrase level and then select the review criteria. Then, it generates cues for participants to reflect on in real-time.

\subsubsection{\textbf{DG2: In-situ citation recommendation as knowledge scaffolding}}

\begin{figure*}[h]
    \centering
    \includegraphics[width=0.7\textwidth]{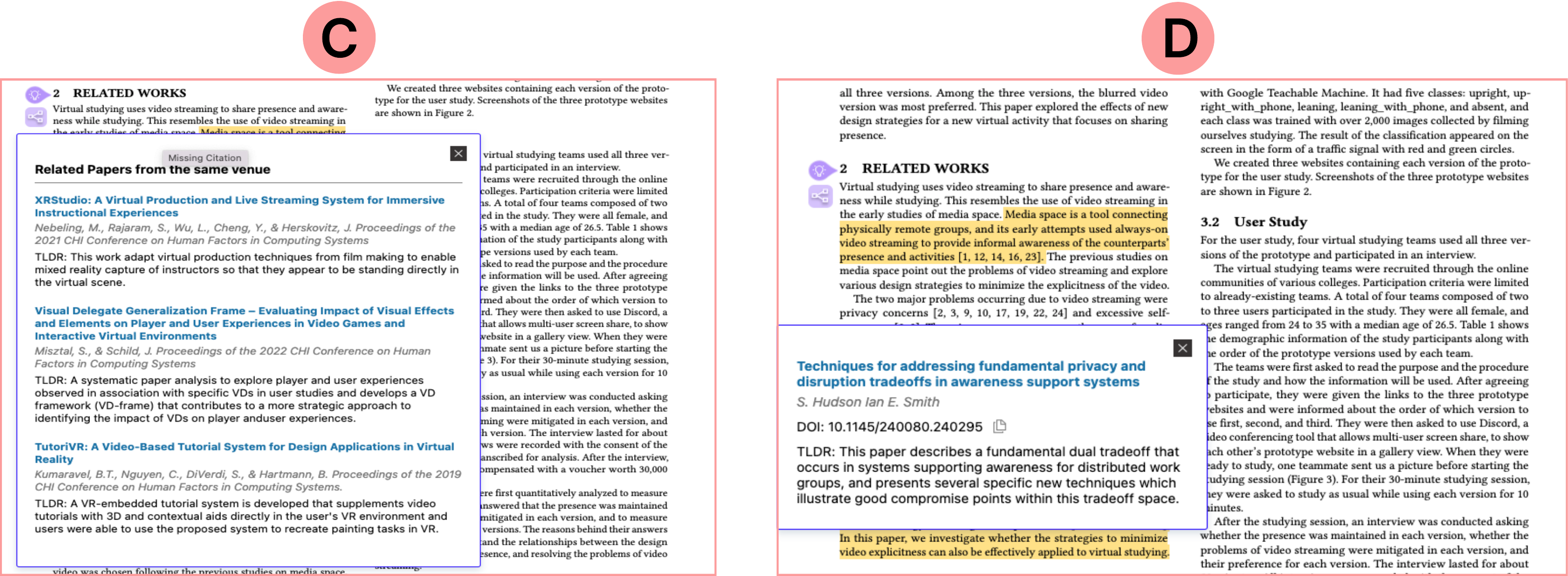}
    \caption{In-situ knowledge scaffolding. ReviewFlow provides (C) a list of relevant papers from the same venue that are not cited and (D) an in-situ citation summary including title, author, and the TLDR summary.}
    \label{fig:in-situ}
\end{figure*}

Prior research in scientific literature review highlights the importance of in-situ knowledge support~\cite{august2022paper,kang2022threddy,rachatasumrit2022citeread,chang2023citesee}. To enable in-situ knowledge scaffolding for assessing novelty compared to prior work, when the user clicks on each reference, ReviewFlow in-situ presents the title and a TLDR summary, as shown in Figure~\ref{fig:in-situ}-D. The TLDR is queried using Semantic Scholar~\cite{semanticscholar}. To raise awareness of the existing references in the reading workflow, ReviewFlow provides a popup window with potential missing citations from the same venues, as shown in Figure~\ref{fig:in-situ}-C.

\subsubsection{\textbf{DG3: Notes-to-outline synthesis}}

\begin{figure*}[h]
    \centering
    \includegraphics[width=0.85\textwidth]{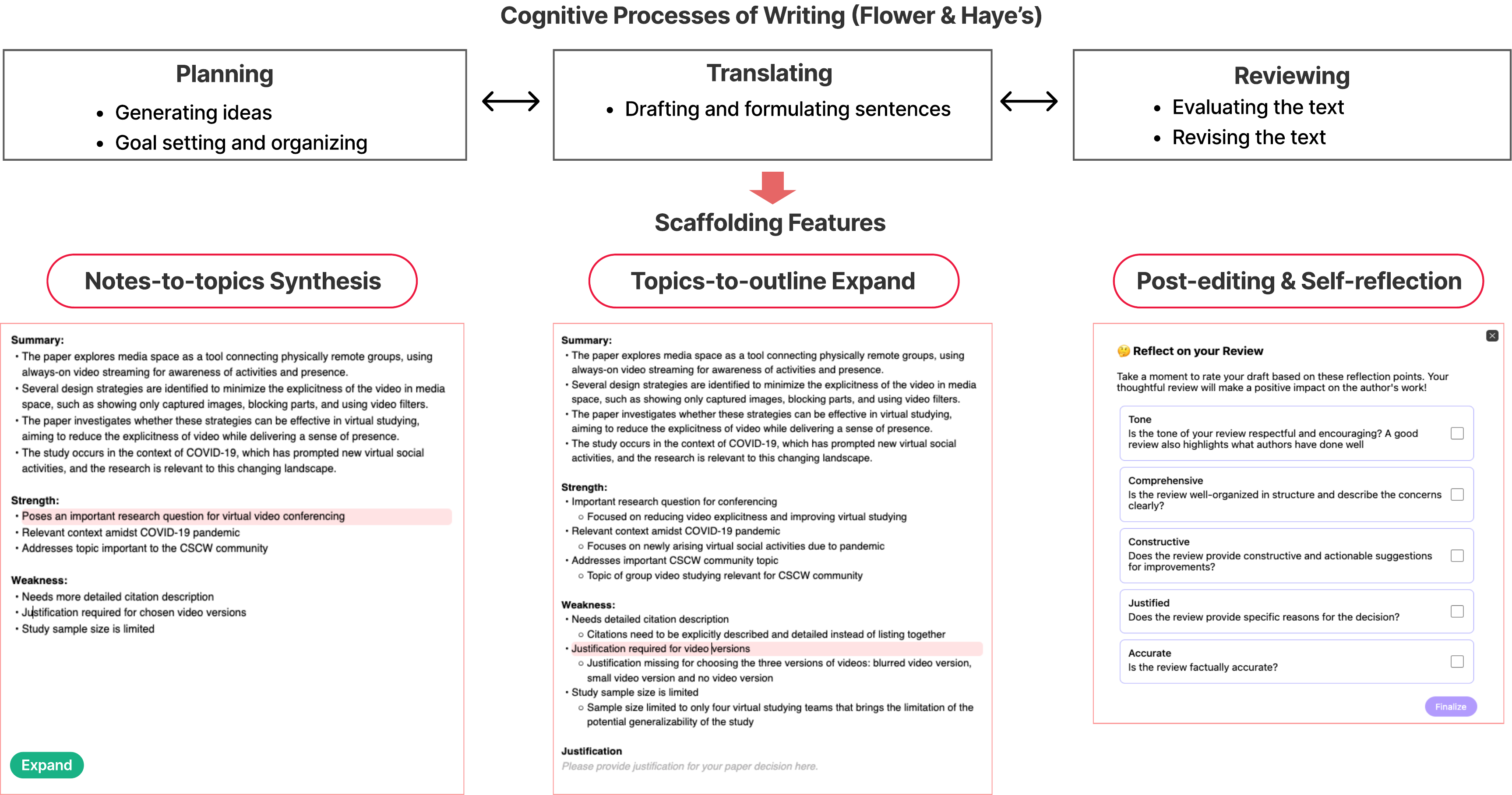}
    \caption{Scaffolding features support the review writing process. The top part shows Flower and Hayes cognitive process of writing~\cite{flower1981cognitive}. The bottom part shows the ReviewFlow features that support each writing step. On the left, ReviewFlow summarized notes into broad topics under strengths and weaknesses to facilitate planning. In the middle, use can click to expand the topics to a detailed outline. On the right, the pop up window encourages self-reflection and post-editing.}
    \label{fig:feature-writing}
\end{figure*}

In the review process, we observed expert reviewers synthesize notes to make a plan on how to draft a review. By modeling the cognitive processes of writing proposed by Flower and Hayes, we designed the \textbf{notes-to-outline synthesis} feature~\cite{flower1981cognitive}. As shown in Figure~\ref{fig:feature-writing}, we decomposed the cognitive processes which include planning, translating, and reviewing, and provided scaffolding features for each stage. In the planning stage, to facilitate writers to generate ideas, set goals, and organize information, ReviewFlow provides a notes-to-topic synthesis that summarizes the user notes together with the highlighted text into general topics, such as ``needs more detailed citation description''. These topics are organized into a structure the aligned with the community practices including summary, strength, and weakness. In the translating stage, to facilitate drafting, ReviewFlow can expand broad topics into detailed bullets if the user clicks the expand button. These bullet points provided a more specific summary based on users' notes. In the reviewing stage, to help users evaluate their written review and revise the text, ReviewFlow pops up a self-reflection card and asks users to self-reflect on their written review using the review criteria, including tone, comprehensive, constructive, justified, and accurate. This encourages users to revise the text if they would like to improve the review quality.

\subsubsection{\textbf{DG4: Fact-checking between outline and source notes and self-reflections}}

\begin{figure*}[h]
    \centering
    \includegraphics[width=0.75\textwidth]{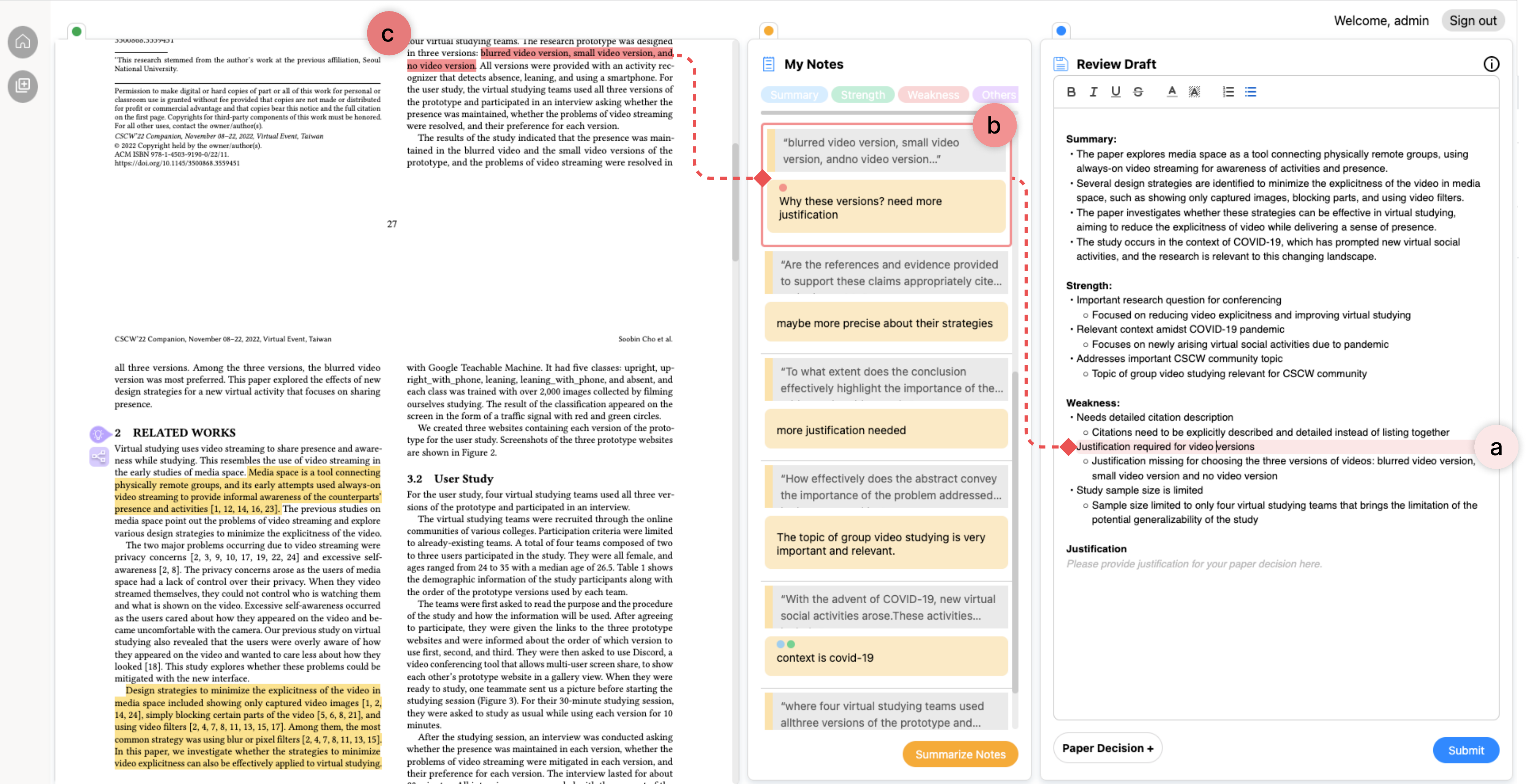}
    \caption{When the participants click on each outline element, ReviewFlow shows the visual mapping among the summarized outline bullet (a), the note in the middle (b) and original pdf with highlights (c). }
    \label{fig:fact-checking}
\end{figure*}

When the research team introduced the idea of using AI to facilitate the reading and writing process, participants expressed the need for fact-checking and providing explanations. To avoid biasing the decision to either accept or reject the paper but encourage justifications, ReviewFlow provided the feature that for each synthesized outline bullet, when the reviewers click it, the notes in the middle column will be highlighted. The PDF on the left column will also scroll automatically and highlight the corresponding location of the notes, as shown in Figure~\ref{fig:fact-checking}. 

% \begin{figure*}[h]
%     \centering
%     \includegraphics[width=0.3\textwidth]{figures/reflection.pdf}
%     \caption{Self-reflection card that asked participants to reflect on their tone and structure of the written review.}
%     \label{fig:reflection}
% \end{figure*}

% To guide an LLM in generating results, various prompting strategies can be employed [57]. Zero-shot prompting involves a questionanswer mechanism, while one-shot prompting provides one example for the AI to base its output on. Few-shot prompting involves providing the AI model with multiple examples, and role prompting requires the AI to complete a task based on a given role rather than examples or templates. Notably, Wei et al. [76] introduced the chain of thought (CoT) to elicit reasoning from the model. This technique enables models to break down complex, multi-step problems into intermediate steps by offering a sequence of prompts that guide them towards a final objective. Implementing CoT significantly enhances the performance of previous language models in reasoning tasks previously considered “off-limits” for language models, even achieving human-parity performance in some tasks [68].

\subsection{System implementation}
ReviewFlow's front-end is a React web application built on top of an existing PDF highlighting library~\footnote{https://github.com/agentcooper/react-pdf-highlighter}. The front end is responsible for displaying the PDF, managing the annotation data, and displaying the text input used to write the review draft. The back-end uses a Flask server which handles the GPT-4~\footnote{https://openai.com/research/gpt-4} API endpoints, MongoDB endpoints for storing data, and GROBID~\cite{grobid} for PDF data extraction. This data includes all of the parsed text of the PDF, its PDF coordinates, and citations linked to their references. Detailed prompts are provided in the Appendix.

\subsubsection{Contextual cues} 
After the user uploads a paper PDF on the front end, a request is sent to ReviewFlow's back-end server where GROBID is used to parse and extract the paper's content, creating an XML/TEI structured document with the coordinates and content of the section titles, text body sentences, and inline citations. This extracted data is then sent to the front-end client where ReviewFlow combines GROBID's section data with PDF.js's section data to make GPT-4 API calls to generate the contextual cues for each of the paper's sections.

Each GPT-4 API call utilizes the section's text as a prompt to generate a contextual reflection question for each of the following critical review aspects: importance, novelty, validity, and clarity. The GPT-4 response is formatted as a JSON and is streamed to the front-end client. To generate phrase-level cues, when the user highlights text on the PDF and clicks on the button with the light bulb icon, a pop-up appears with the review criteria. When the user selects one of the aspects and clicks "Get a Question'', GPT-4 will use the selected aspect, highlighted phrases, paragraph of the highlighted phrases, and paper abstract to generate cues.

\subsubsection{In-situ citation recommendation}
When the PDF is initialized, ReviewFlow uses GROBID's XML/TEI data to create a citation layer that overlays all in-line citations. When the user clicks on an in-text citation, a popup shows the paper title, publication date, DOI link, and a short description that is searched using Semantic Scholar API~\cite{semanticscholar}. To recommend corresponding papers, the system calls the Semantic Scholar Recommendation API\footnote{https://api.semanticscholar.org/api-docs/recommendations} and uses the keywords of the paper together with the venue to retrieve the most similar paper. After filtering out the retrieved papers that have already been included in the current paper, the top three papers are added as a citation pop-up.
% This citation layer is constructed by aligning GROBID's scaled coordinates to the client PDF's coordinates to create an HTML division element that overlays the in-line citation. This new element is bound with an event listener such that when the user clicks on the element, a popup appears with the corresponding citation data. The citation data includes GROBID's citation data such as the title, publication date, and DOI. 
% Additionally, a Semantic Scholar API call is made for each citation to generate a TLDR description of each paper and to populate any missing fields that are not already provided by GROBID. 

\subsubsection{Notes-to-outline Synthesis}
After the user creates multiple notes, a ``Summarize Notes'' button will appear at the bottom of the notes panel. Upon clicking this button, the front-end client will organize all the user's notes data including the corresponding highlighted paper content, selected tags, note text, and paper abstract. Subsequently, the system will use this data as prompt. We use the few-shot learning paradigm with the corresponding prompts~\cite{liu2023pre}: ``Please create three important topics on the paper's [strengths] and [weaknesses] with less than ten words that combine and summarize the user notes.'' The output is then formatted in JSON and streamed directly into the front-end's text-editable draft panel with a summary and topics organized on strengths and weaknesses. When the user clicks on the ``Expand'' button at the bottom of the draft panel, all of the text in the draft text input box, notes, and paper abstract is sent to the back-end server and requested a GPT-4 API call. The GPT-4 API call will then respond with a streamed JSON-formatted output with more details based on the user's notes and topics specifically for the strength and weakness sections.

% : ``Please create three important topics on the paper's [strengths] and [weaknesses] with less than ten words that combine and summarize the user notes.''
\section{Method}
We designed a within-subjects experiment to answer the questions below:
\begin{itemize}[leftmargin=*]
    \item RQ1: How does ReviewFlow affect participants' final written review quality, compared to the same system with no intelligent scaffolding?
    \item RQ2: How does ReviewFlow affect participants' workflow, in terms of time and engagement?
    \item RQ3: What benefits and challenges do participants perceive with ReviewFlow's intelligent scaffolding? 
\end{itemize}

\subsection{Study Design}

We conducted a within-subjects experiment with 16 novice participants where each participant experienced both the ReviewFlow condition and the Baseline condition in two sessions separately. We counterbalanced the order of two conditions and papers using a Latin Square design to minimize the potential order effects. To reduce knowledge transfer and minimize fatigue, we scheduled the two study sessions of each participant at least 24 hours apart. The ReviewFlow condition includes all scaffolding features, while the Baseline condition includes the minimal guideline. In the baseline interface, users can still highlight, take notes, and tag notes. Community guideline is also provided.
% We choose between subject experiments instead of within-subjects because (1) we want to control the study duration considering participants' energy levels. The average reviewing time reported by experts for a whole long paper is 4.75 hours, in the observational study, experts took about 45-60 minutes to review one short 2-4 pages paper. (2) we are recruiting novice reviewers, having within-subject experiments may lead to knowledge transfer between two conditions; However, this brings into some random noise in the study, e.g., participants' stress level and energy of that day.

% We randomly assign each participant to two conditions: an experiment condition where participants received all scaffolding in the peer review process; versus a control condition where participants did not receive additional scaffolding. This comparison can provide insights into user's performance and reactions of scaffolding in different writing stages. We simulated a real peer review process where participants play the role of a reviewer with the task of writing a peer review. We counterbalanced the order of each condition through random assignment. 

\subsubsection{Participants}

Before the study, we sent out a pre-study survey to participants to collect information about their expertise, research topics, review experience, knowledge level on AI, and demographics (e.g. age, gender, race). We recruited 16 participants who have experience in conducting academic research for at least two years and have zero to twice conference peer review experience to make sure that they are novice reviewers who lack review experience but are equipped with domain knowledge. We advertised recruitment messages to colleagues, mailing lists, communication channels, and social media and recruited participants from four universities across the US. Using a snowball sampling approach, we asked participants to refer their friends and colleagues. Participants have on average 1.2 years of writing and submitting academic papers. Each study section takes around 90 minutes and all participants were compensated for \$20 per hour. The study is IRB-approved.

\subsubsection{Procedure}

Before the user study, participants filled out a pre-survey that captured their previous experience in reviewing and reviewing papers on HCI/CSCW conferences and their expertise in HCI/CSCW.  We asked participants whether they had read the paper before to make sure they all were seeing the work for the first time. Participants conducted reviews on two papers using different versions of our interface (ReviewFlow or Baseline). To counterbalance the order effect, we randomized the order of the control condition and the experiment condition for each participant, so half participants encountered ReviewFlow in their first session, and the other half experienced it in their second session. For each session, we followed the review process used at the CSCW conference, where we provided the paper draft and the review guidelines. 

For both sessions, participants were told to spend around 60 minutes -- and no less than 30 minutes -- on the review, but they could take as much time as needed. In the pilot study with 2 participants who had very little review experience, we found that participants could finish reviewing within 45 minutes. Before each condition session, we gave the participants a quick 2-minute demo of the interface, while in the experiment session demo, we also introduced the functionality of ReviewFlow. After each session, the research team asked the participants to fill out a post-survey to evaluate the system and assess their self-efficacy. After the second session, we conducted a 15-minute semi-structured post-interview to ask open-ended questions about their experience, perceptions, and feedback. The post-interviews were video recorded with participants' permission and were transcribed into text for later analysis. 

\subsubsection{Paper Selection Process}
% \footnote{https://openreview.net/forum?id=Hye5TaVtDH}\footnote{https://openreview.net/forum?id=B1xxAJHFwS}.
We collected recent one-year papers from HCI-related conferences including CHI, CSCW, UIST, UBICOMP, IUI, DIS using the list provided by \footnote{http://www.conferenceranks.com/}. To make our study time short so that people have the energy to finish the task, we filtered papers that had fewer than 3000 words and further filtered out papers that had technical terms and jargon. The two papers we selected need to have similar lengths, similar difficulties, and from the same conference venues. Combining all the criteria above, two papers from CSCW Companion were selected for the study. The first paper includes the keywords ``virtual studying, video streaming, awareness'' and contains 2853 words. The second paper includes the keywords ``virtual environment, cross-lingual collaboration, team formation'' and contains 2910 words. 

\subsubsection{Measures}
We collected a mix of quantitative and qualitative data, including each participant's log data that captured their interactive behaviors with the system, the final review written for each paper (N=32), the post-survey, and the interview transcripts. The research team analyzed these combined sources of data to reveal insights.

\paragraph{Quality Ratings on the Final Peer Review }

To measure the quality of the review, we recruited two experts who have conducted research for more than three years of review experience in HCI or CSCW conferences. They counted the number of strengths and weakness in the review (a proxy for coverage) and rated the quality of all final reviews (N=32) with a five-dimension rubric based on reviewer guidelines for CSCW conference and previous research~\cite{yuan2021can}:
\begin{itemize}
    \item {Tone}: The tone of a peer review is always encouraging and respectful. A good review also highlights what authors have done well.
    \item {Comprehensive}: A good review is always well-organized in structure, which includes a summary, strengths, weaknesses, and a clear description of concerns.
    \item {Constructive}: A good review usually provides constructive suggestions. Following the weakness, reviewers usually will provide actionable items that the author can work on to improve the paper's quality.
    \item {Justified}: A good review justifies specific reasons for their decisions. Avoid providing a decision without any supporting evidence.
\end{itemize}

Two experts rated these five dimensions on a simple seven-point scale (1-7). Each expert first read one paper and wrote a peer review of the paper. After this process, the research team provided provided instructions and two examples for them to rate and discuss until they reached a consensus on ratings based on the instructions. Then, they rated each dimension of the review on the paper independently. The inter-rater reliability between two experts on all 32 data is moderate where Krippendorff's alpha is higher than 0.50~\cite{krippendorff2018content}. The research team then used the average scores by two experts for each dimension. 

To construct the proxy of review quality, we further asked the experts to count the number of strengths and weaknesses raised by the participants in each review. Another proxy of the review quality we used is the participants' self-rated satisfaction with their reviews. After participants submitted the reviews, we asked them to rate their satisfaction with their reviews on a scale of 1 -7 (1 as not satisfied at all and 7 as very satisfied).

\paragraph{User Interaction Data}
To measure participants' interaction with the tool, we instrumented the interface to collect a range of user activity log data. We collected two timing measures -- how long each participant took to finish the review session and how long each participant spent editing the review within the text box. The entire review session time includes reading time and writing time.  
The ReviewFlow interface also collected interaction data to capture how much each participant interacted with each scaffolding feature, including the times they answered the contextual cues, clicked on the citation summary, checked the citation pop-ups and used the outline summarization feature. 

\paragraph{User Preferences and Reactions} 

To evaluate users' preferences for the ReviewFlow experience compared to a baseline plain review editor, we asked participants to fill out a short post-study survey. The survey asked participants to directly compare the perceived usefulness, enjoyment, easiness, and sense of control between the ReviewFlow and baseline system. We collected the level of cognitive demand using NASA TLX on a scale of 1-5~\cite{hart1988development}. 
We further collected the feeling of control and collaboration on a scale of 1-5.
Then we specially asked participants to evaluate each of the scaffolding features. The survey collected their 5-point Likert scale ratings on perceived usefulness, and perceived accuracy for each feature. 

Previous research showed that scaffolding can promote learners' self-efficacy~\cite{yantraprakorn2013enhancing}. Here, we measured whether using the scaffold system can improve novice reviewers’ self-efficacy and confidence. We asked each participant to report self-efficacy after using the ReviewFlow system and the Baseline system by answering the question ``How confident are you in your ability to write a conference peer review next time after using the system (1 as not confident and 7 as very confident)''.

After the post-survey, we conducted a 15-minute semi-structured interview with all participants to capture their overall thoughts as well as specific perceptions of machine-generated highlights and summaries. For example, the research team asked ``What do you think of the difference between the task with and without the support of ReviewFlow'', ``What did you learn after writing the review with the system'', and ``What concerns did you have when using ReviewFlow?''.

\paragraph{Users' knowledge of each paper's topic as control variables}
We measured participants' existing knowledge of the two papers as a control variable. Participants described their knowledge of each paper's topic from not familiar as 1 to very familiar as 7 in the post-survey. All participants reported that they had never read and remembered the papers before. The average rating of knowledge level on the first paper is 3.46 and the average rating for the second paper is 3.42.
% std as 2.24, std as 2.18
% On the post-survey, participants self-reported their knowledge of each paper's topic from 1=not familiar to 4=very familiar. Participants' average familiarity was 2.1 with a standard deviation of 1.4. The familiarity was slightly higher for one of the paper topics than the other, so we included this as a control variable in our statistical analyses comparing the conditions. All participants reported that they had never read or remembered the specific papers before.

\subsection{Analysis}
% To explore these research questions, we collected quantitative data through pre-study surveys, system logs, including session time and paper writing time, and their final decision along with the written review. We also gathered qualitative data through post-study surveys and semi-structured post-interviews.

\paragraph{Quantitative data analysis}
To measure the effect of the ReviewFlow system on each dimension of the review quality (eg. tone, comprehensive, constructive and justified), we conducted repeated measure ANCOVA tests. We used paper ID, the order of experiment conditions (whether ReviewFlow was used for the participant's first or second paper), the knowledge level of each paper topic as co-variants. To measure the effects of the experiment condition on the time they spent reading independent reviews and writing reviews taking the consideration the differences between the paper topic and order effects, we ran a repeated measure ANCOVA using the paper ID, the order of experiment condition and the knowledge level of each paper topic and the review word counts as co-variants. To compare participants' behaviors between the two conditions, we conducted Wilcoxon signed-rank test, which is a non-parametric rank test, on users' interaction data, e.g. number of notes participant taken in both conditions~\cite{woolson2007wilcoxon}.

\paragraph{Qualitative data analysis}
All semi-structured interviews with participants are recorded and transcribed. Two researchers conducted iterative open coding on the transcripts using Dovetail\footnote{https://dovetailapp.com/} and conducted thematic analysis on the transcripts~\cite{braun2006using}.  They open-coded the data by identifying topics mentioned by the participants. Initial codes were combined into preliminary themes, which were discussed among the research team. After iteratively discussing the code themes, researchers identified the final themes around: participants' reactions to each feature, their overall perceptions of the ReviewFlow system and their reactions to each scaffolding feature.
\section{Results}
We report our results from the within-subjects experiment and the post-interview. In the within-subject experiment, across both conditions, participants spent an average of 38.7 minutes writing a review with an average length of 243 words written. 60\% of the participants decided to accept the paper, which is consistent with the original decisions for the two papers. Our findings revealed that ReviewFlow provided more guidance to novice reviewers and made the review process more useful and engaging.
% However, participants surfaced their potential concerns about using AI in the conference peer review process.

\subsection{RQ1: ReviewFlow helped participants write more comprehensive reviews} 

\begin{figure}[htb]
    \centering
    \includegraphics[width=1.05\linewidth]{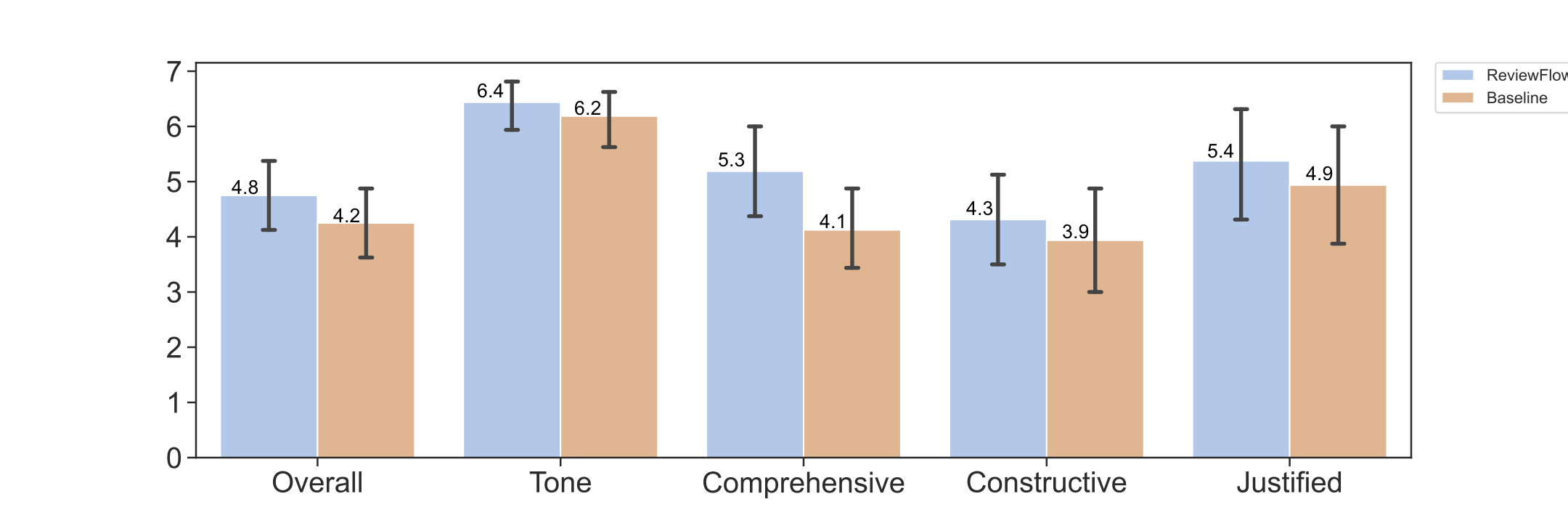}
    \caption{Experts' evaluation scores on the quality of review written by ReviewFlow versus Baseline.}
    \label{fig:expert-eval}
\end{figure}

To compare the quality of written review in both conditions, we performed ANCOVA tests to examine the effect of the two conditions on each quality measure, accounting for the review length, participants' knowledge of the topic, the order of the experiment condition as co-variants.
As shown in Figure~\ref{fig:expert-eval}, we found that the reviews in the ReviewFlow condition are significantly more comprehensive than the Baseline condition (p = 0.04$*$, F = 3.55).  We found no significant interaction effect between the order of the experiment conditions and no statistically significant interaction effect between the knowledge of the two papers. 

Furthermore, we evaluated several proxies of review quality including the number of strengths and weaknesses and participants' self-rated satisfaction with the written review. 
As shown in Table~\ref{tab:performance-quality}, participants wrote longer reviews and called out more strengths and weaknesses in the paper in the ReviewFlow condition. However, there is no significant improvement in the constructiveness of the review. This indicates that the ReviewFlow system can scaffold participants to capture more pros and cons but still cannot help participants write constructive solutions for each weakness. Also, participants' self-rated satisfaction with reviews written using the ReviewFlow tends to be higher than using the Baseline. ANCOVA test with proxies did not a show significant effect.  Participants reflected on the reason that ``I captured more pros and cons since I took more notes this time, so I feel satisfied with the review as it covered more aspects'' [P3].

\begin{table}[htb]
    \centering
    \begin{tabular}{l  r | r  | c | c}
          \multicolumn{2}{r|}{ReviewFlow} & Baseline  & p & F \\ \hline
          \multicolumn{2}{l|}{\textit{\textbf{Proxy of review quality}}}   & & \\
          Count of strengths & 2.38 (0.19) & 1.91 (0.17)&  - & -\\
          Count of weaknesses & 2.62 (0.22)) &  2.08 (0.27) & - & -\\
          Self-rated satisfaction & 4.35 (0.4) & 4.00 (0.34) & - & - \\ \hline
         \multicolumn{2}{l|}{\textit{\textbf{Length of the review}}} &  & & \\
          Word counts& 250.8 (19.4) & 235.8 (23.5) & - & - \\ \hline
          \multicolumn{2}{l|}{\textit{\textbf{Time on task}}} & & & \\ 
            Reading time (minutes)  & 26.9 (1.9)& 19.2 (2.1) & *** & 19.2 \\
            Writing time (minutes) &  15.5 (2.2) & 15.7 (2.4) & - & - \\ \hline
          \multicolumn{2}{l|}{\textit{\textbf{Self-efficacy}} } & & &  \\ 
          Self-efficacy on reviewing & 4.92 (1.43) & 3.92 (1.36) & * & 4.2 \\
    \end{tabular}
    \caption{Proxies of review quality include the number of strengths, number of weaknesses, and satisfaction on the written review rated by participants themselves. Participants included more weaknesses in the ReviewFlow condition. Participants spent more time reading and reported higher self-efficacy after using the ReviewFlow.}
    \vspace{-2em}
    \label{tab:performance-quality}
\end{table}

\subsection{RQ2: Longer interaction duration with ReviewFlow but improved participants' self-efficacy} 
\subsubsection{Participants took a longer time to read papers with ReviewFlow but a shorter time on drafting the final review}
We performed ANCOVA to examine the effect of the two conditions on writing time, accounting for the length of the review, participants' knowledge of the topic, and the order of the experiment condition as covariates. As shown in Table~\ref{tab:performance-quality}, participants spent significantly more time in reading and sense-making with the ReviewFlow than with the Baseline. However, they spent less time on writing, while the result was not significant. 68.8\% of them still reported that ``ReviewFlow saved me more time on writing reviews''. P11 reflected the reasons that ``Answering pop-up questions made me spend more time on reading the paper, judging it from different aspects, and taking notes, but I feel it did save me time in the end since I don't need to go through all the notes again''[P11]. Similarly, P6 also mentioned the reason for saving time as ``having that[my notes] summarized and organized at the end meant that I didn't have to go back through each section and think about where I should put the strengths and weaknesses''[P6]. On the contrary, P5 reflected that they spent more time considering different aspects, such as validity, novelty, clarity, etc. P5 highlighted that ``I can write the review fast, but that is not my goal. As a reviewer, it is more important to spend enough time carefully evaluating the paper''[P5].

\subsubsection{Participants reported to have higher self-efficacy after experiencing ReviewFlow than the Baseline}
We asked participants to rate their self-efficacy on the ability to conduct a conference peer review and conducted the ANCOVA test between each participant's ratings, accounting for covariates. Table~\ref{tab:performance-quality} showed that self-efficacy ratings in the ReviewFlow are significantly higher than the Baseline. This result indicated that participants built more confidence in learning after using the ReviewFlow. Specifically, participants described their learning process --``I feel I gradually built confidence. At the beginning, I would try to answer every question. But later on, I started to remember which aspect I should think about while reading that section, so I didn't need to frequently check guided questions''[P13].

\subsubsection{Participants took more notes in the ReviewFlow and used the features actively}

Participants took significantly more notes in the ReviewFlow condition (M=12.9, STD = 3.7) than in the Baseline condition (M = 9.1, STD = 3.9, Wilcoxon signed-rank test, Z= 103.0, p <= 0.01, $**$). This indicates that participants are more engaged in the reading process with ReviewFlow and not only reading text but also critically reflecting on the content. Participants actively answered the guided questions. All participants used the notes-to-outline features which helped them summarize these notes into an outline. Three participants did not expand the high-level outline to a detailed outline.

\subsection{RQ3: Participants perceived ReviewFlow as useful in the review workflow} 

\begin{figure*}[h]
    \centering
    \includegraphics[width=0.7\linewidth]{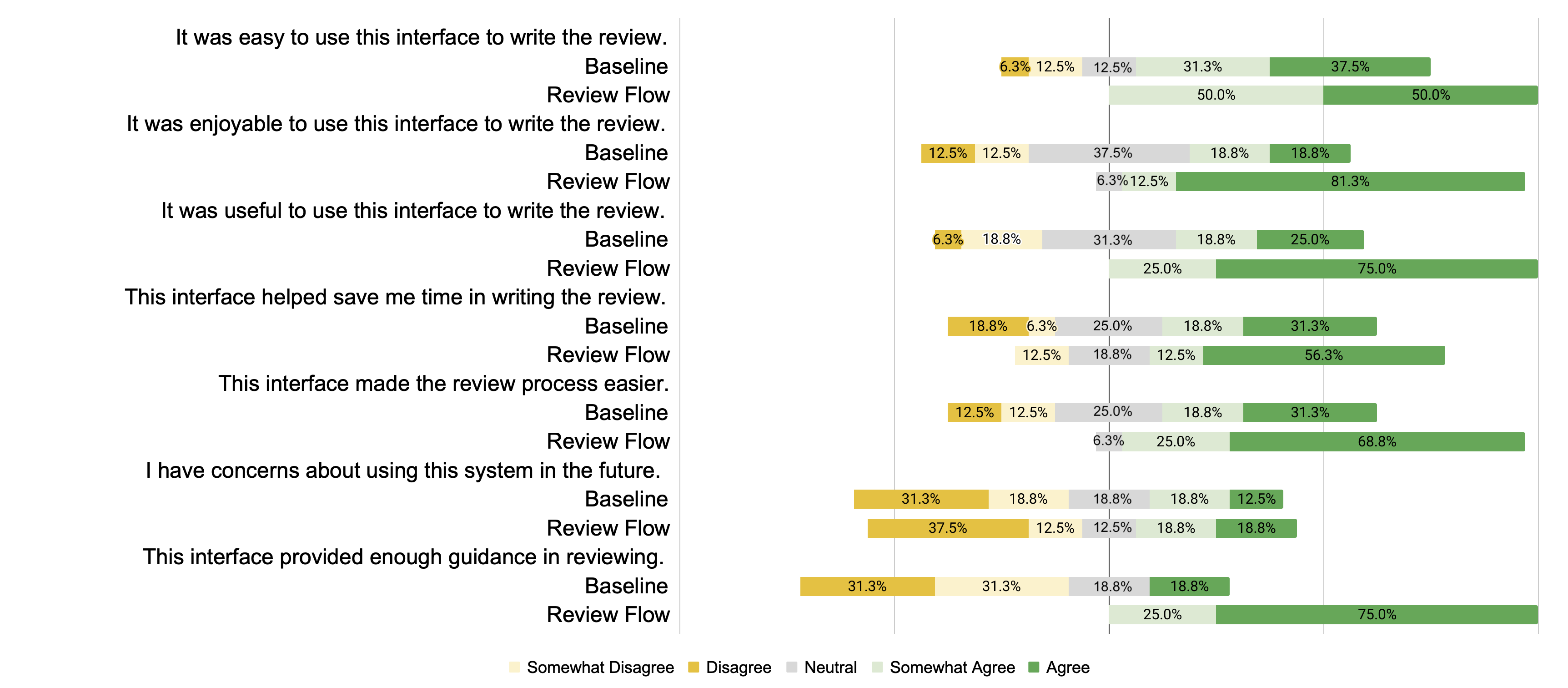}
    \caption{Participants' reactions to the system in two conditions in terms of easiness, enjoyment, and efficiency.}
    \label{fig:interface-compare}
\end{figure*}

After participants had experienced the two systems, we asked them to compare two conditions. As shown in Figure \ref{fig:interface-compare}, participants highly preferred ReviewFlow over the Baseline. 93.7\% of participants perceived the interface as enjoyable to use. Participants mentioned that the scaffolding features make the review process more engaging and less boring (N=5). All participants think that the system is useful and they like the guidance provided in reviewing. For instance, P11 described the guidance they received as similar to experts -- ``I feel like some experts, such as my advisor, were sitting next to me to prompt me in particular ways and show me how to write the review''[P11].

\begin{figure*}[h]
    \centering
    \includegraphics[width=0.7\linewidth]{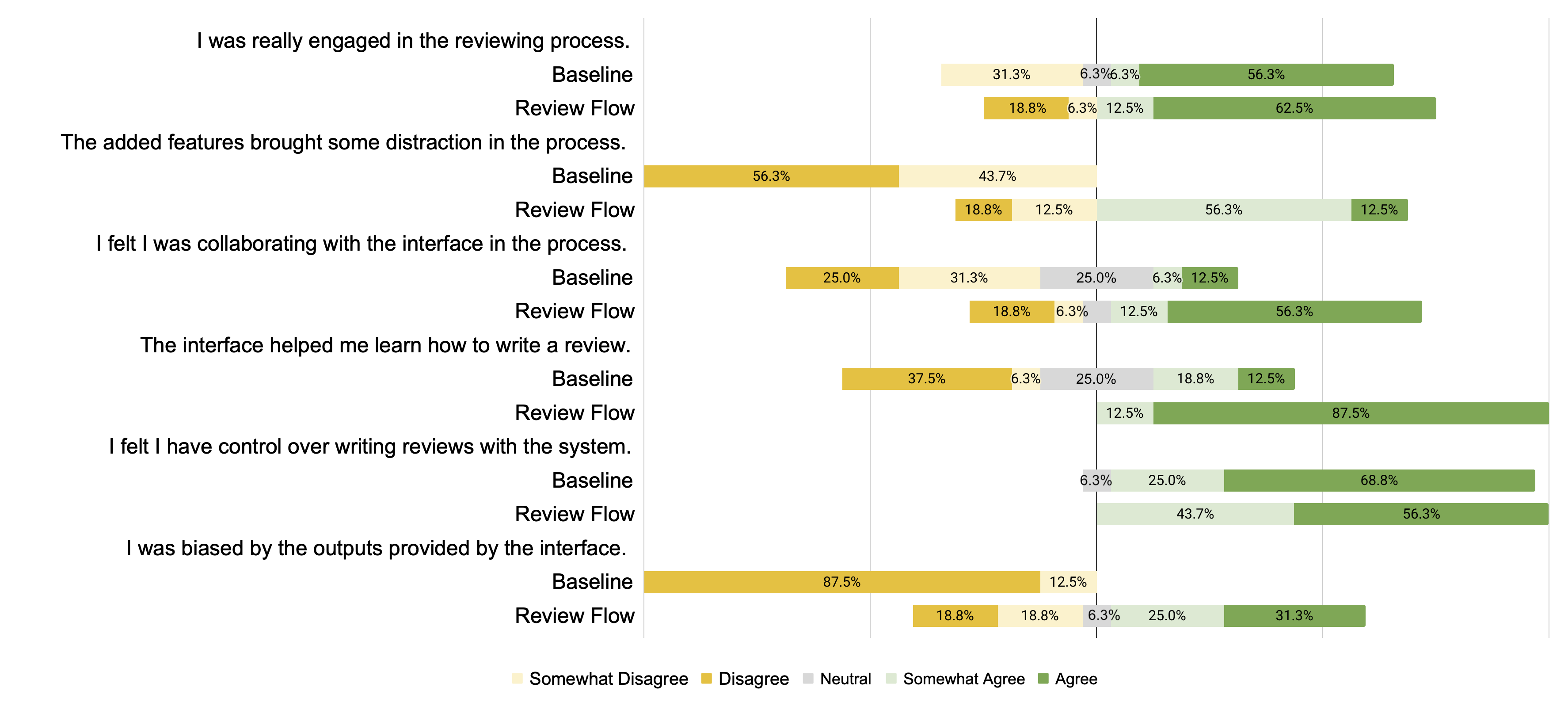}
    \caption{Participants' feelings of the engagement, control, and perceptions of bias in two conditions.}
    \label{fig:interface-perception}
\end{figure*}

 % ~\cite{hart1988development}~\cite{wu2022ai}
We further measured participants' perceptions of the tasks in each condition. In the post-survey, we asked participants to rate cognitive workload, such as distraction and engagement, the feeling of control and collaboration with the interface, and perceived learning gains. As shown in Figure~\ref{fig:interface-perception}, 75\% of the participants reported that they are engaged in the process, but the add-on scaffolding features bring in some distraction for 68.8\% of the participants. However, participants highlighted that it is not a bad distraction, but served as a staging process that motivated them to think. For instance, P8 explained that ``Since you don't want to just keep going through the paper without getting anything from it. The distraction is like a pop-up that keeps prompting you about review criteria from different places''[P8]. 

All participants believe that the ReviewFlow helped them learn the peer review process. 68.8\% of participants have the feeling of collaborating with the interface. Participants identified the collaboration mainly happened during the process of answering contextual cues in each section and using the note-to-outline synthesis feature. The two-step process where participants can first summarize notes into a high-level outline and then expand it provided them the feeling of ``iterating with an assistant''[P9]. Even though the current system uses AI-generated outlines, all participants agreed that they still have the control over writing process. One participant mentioned the reason is that ``It just synthesized what I have already written in the notes and I can still write it by myself''[P9].
37.5\% of the participants described that the notes synthesis did not bias them in the current session, but they still worried that ``busy reviewers might randomly create notes and use the outline to make a decision''[P14].

\subsubsection{Participants have different preferences on the scaffolding strategies}

As shown in Table~\ref{tab:feature-usage}, participants perceived that section-level cues were more useful and more accurate than the phrase-level cues on the highlighted text. Two participants reflected that the uncertainty on the phrase-level cues is high since the questions' quality depends on the number of words they highlighted. P15 also said ``I have already had a question in my mind when I highlighted certain parts of the text. So if the question did not match with the question I was thinking, I felt a bit disappointed''[P15].  
We also found that many people did not pay attention to the missing citation support, since they did not focus on evaluating the related work section. 
The most useful feature perceived by participants is the notes-to-outline synthesis feature where participants reflected that it helped them ``easily get a sense on top of my notes''[P3]. However, three participants did not expand the topic to a detailed outline. They reflected that the detailed outline bullets contained similar content as the high-level outline topic which made it less useful. We further conducted an in-depth analysis of how participants used and perceived different types of scaffolding strategies:

\begin{table}[htb]
    \centering
    \small
    \begin{tabular}{ p{0.5\linewidth} | c | c  |c |c }
          & \#(p) & freq& useful &  accuracy \\ \hline
        Section-level cues &  14 & 3.8 & 5.4& 4.9\\ 
        phrase-level cues & 10 & 1.6& 3.8 & 4.3\\ 
        In-situ citation pop-up & 10 & 1.0& 4.6 & 6.2\\
        Missing citation pop-up & 9 & 1.0& 2.5 & 3.4\\
        Summarize notes to high-level topics & 16& 1.12 & 6.3 & 5.2\\
        Expanded topics to detailed outline& 13& 1 & 5.2 & 4.3\\
    \end{tabular}
    \caption{Usage of scaffolding features shows the number of participants, average frequency, and participants' rating on the usefulness and accuracy. All participants used the features that synthesize notes to high-level topics, while three participants did not expand it to a detailed outline.}
    \label{tab:feature-usage}
\end{table}

\paragraph{Contextual cues guided novices to evaluate from different criteria}
Participants described that the contextual questions are ``more specific and tailored to the paper compared with the general community instruction''(N=3). Instead of providing the review criteria with general questions, such as ``how well does the paper execute their contribution?'', participants preferred to have more contextually guided questions according to the paper content, such as ``does the method section provide clear and well-justified explanations of the research prototype's design including the choice of videos and the method to set up the activity recognizer?''. Participants described that their process of using the section-level cues is slightly different from the phrase-level cues. P10 described that ``I quickly checked out the questions next to each section before I dived into the reading. After I finished reading the section, I came back to these questions and reflected on how I felt''[P10]. Correspondingly, phrase-level questions are used when ``I feel confused or not sure what I should think about''[P11]. These contextual cues reminded them or prompted them to evaluate different aspects of each section.

\paragraph{Notes-to-outline synthesis helped participants structure their notes according to community practices}
All participants used the notes-to-outline synthesis. Participants preferred having notes in a ``structured version'' and indicated that they didn't need to summarize and map their notes by hand. The ability to summarize notes into each review section helped them "reflect on the strengths and weaknesses" (N=2). As a result, they perceived that the outline generation feature saved their time in writing (N=5). Participants liked the fact-checking function that when they can map each topic back to their notes and original pdf, as shown in Figure~\ref{fig:fact-checking}. The visual mapping helped them cross-check the notes and increased their trust in the topic. However, three participants intentionally avoided using the expand button to get a more detailed bullet. They are concerned that the expanded outline may generate content from nowhere. P3 explained that ``I was worried that if I clicked to expand the themes into an outline, it would have incorporated things that I didn't want to include''[P3]. 
% Getting back to the highlight help them validate the source

\subsubsection{Participants' concerns about using ReviewFlow}
Participants are mostly concerned about the potential errors that AI can make in the notes synthesis process. Participants did not express severe concerns in the study session since the generated outline mainly reused or summarized their notes. However, they were still worried that the nuanced tone in the outline produced by AI may exaggerate the weakness of the paper and influence reviewers' perceptions. For instance, one participant mentioned that if you wrote the notes as ``the details around study participants seems a little bit unclear'', but the AI turns it into ``this paper lack clarity'', which may mislead the reviewer. Moreover, participants expressed their concerns of other reviewers in the community. They are worried that ``last minute reviewers may randomly leave notes and then use the tool to generate an outline. If they expand the notes it might not be truly reflective of what they thought if they were not paying attention. The misuse of the tool will not be fair to the paper authors and the research community in general''[P14].

\section{Discussion}
Advances in large language models are changing how work gets done. Our research explores how we might integrate intelligent scaffolding in a way that informs and guides novices and steers away from biasing the underlying judgment. 
To inform the design of intelligent scaffolding, we conducted a formative study where we learned what strategies and practices experts adopt, as well as, what challenges novices face when writing peer reviews.  Our ReviewFlow system aimed to leverage LLMs to provide timely considerations while reading and assessing a submission and structural support when composing a review.
Our within-subject experiment (N=16 participants) found that novice reviewers not only preferred ReviewFlow over the baseline system, but they also wrote longer and more comprehensive reviews, as judged by experts. Using ReviewFlow participants spent more time reading the paper and took more detailed notes. Those notes were aided by ReviewFlow's contextual cues, and helped participants call out more strengths and weaknesses. Using ReviewFlow, participants were more satisfied with their reviews (according to self-ratings) and attributed this to the timely cues and a useful workflow.  

% A more in-depth comparison with related works and a discussion on how ReviewFlow aligns with or diverges from current practices in academic peer reviewing would provide a richer context for understanding the tool's significance.
% 40\% of participants chose to reject decisions, which is the opposite with the original decisions for the two papers. 
% Engaging longer time on reading and reflecting
% discuss the constructiveness? --> current scaffolding strategy did not contribute to a more constructive review.
In comparison with general peer review practices~\cite{wang_reviewrobot_2020,shah2022overview}, participants using ReviewFlow allocated more time to reflect on each section and evaluate the paper's quality. Interestingly, 40\% of participants opted for rejection decisions in both conditions, contrary to the original decisions for the two papers. A typical peer review not only contains the paper summary and its contribution but also raises weaknesses from different aspects together with constructive feedback or thought-provoking questions~\cite{moore2013critical}. However, participants in both conditions still found it hard to provide constructive feedback on each weakness,  even with example reviews. Perhaps future systems could incorporate additional features into the intelligence scaffolding, such as presenting examples of improvement suggestions from similar prior reviews or offering feedback to the reviewers after they have an initial draft~\cite{krause2017critique}.

\subsection{What role should machine intelligence play in this review process?}

AI has come a long way since the early and annoying attempts at supporting work (e.g. Clippy). With the development of LLMs, AI can automate or augment many aspects of knowledge work, including information discovery, sensemaking, and writing~\cite{palani2023relatedly,liu2023selenite,zhang2023visar}. 
While LLMs have become incredibly valuable, the risk now might the risk now might be the tendency, especially among tech-focused innovators, of taking automation too far and dehumanizing work. Prior research has explored whether models can predict a paper decision or even draft a review~\cite{liang2023large}. Our ReviewFlow system was designed with the intention of balancing the use of technology with human values. Our system strikes this balance by taking cues from the learning sciences research on scaffolding~\cite{saye2002scaffolding,holton2006scaffolding}. The goal of ReviewFlow is not to produce reviews, per se, but to convey an understanding of how to think while reading a submission and writing a review.  Our values place more emphasis on training and preparing novices (perhaps for their next peer review), not just on getting to a final peer review. 

%Further discussion on scaffolding. In the peer review process, the current study tends to scaffold novices. what about experienced reviewers.
Prior work indicates that well-design scaffolding can help novices operate more like experts~\cite{yuan2016almost}. The concept of Zone of Proximal Development (ZPD) represents the space between what a learner is capable of doing unsupported and what the learner cannot do even with support~\cite{harland2003vygotsky,wass2011scaffolding}. Scaffolding works most effectively when it meets novices when they need support and tapering off when they have internalized the best practices.
The scaffolding in ReviewFlow is intelligent and contextual. For instance, users get reflection cues that adapt specifically to the current section. 

ReviewFlow was explicitly designed to support novices, but it does not adapt to someone's existing knowledge or experience with the task. A longitudinal deployment would provide insight on whether novices continue to prefer ReviewFlow, or whether it is useful for early experiences with peer reviewing. As novices become more proficient, they may eventually prefer to work with less structure (e.g. traditional text editor), although even old timers likely find value in ReviewFlow's pragmatic support for capturing and synthesizing notes. Future research could extend the current system to be more context-aware and more adaptive to users and then study its use with both novices and experienced reviewers over time to gain design insights for interactive scaffolding. Furthermore, future work could explore other properties of scaffolding, such as the timing of when it's provided, the communication modality, the format or representations used, or the strategies leveraged, such as comparing generated examples vs. guiding questions.

% AI scaffolding for peer review writing
    % We need ablation studies: compare each type of scaffold each time
    % More variations of scaffolds: examples versus guided questions
    % Different community guidelines or domains may have different workflows, but the scaffold intervention still works
ReviewFlow incorporates three types of scaffolding to support cognitive activities: contextual scaffolding generates reflection questions to aid in paper reading; knowledge scaffolding focuses on citations to facilitate paper evaluation; and structural scaffolding synthesizes notes into structured outlines to assist in review writing. The current study provided insights into user perceptions of different scaffolding by observing their usage of all the features and asking them to rate each feature's usefulness and accuracy. 
More rigorous and well-controlled ablation studies could help us understand the underlying factors impacting the ReviewFlow experience. For instance, a study with three experimental conditions, each providing only one type of scaffolding---contextual scaffolding, knowledge scaffolding, and structural scaffolding---would allow for a more comprehensive evaluation of the quality of review writing, task efficiency, and the learning impacts for the user.

\subsection{How can intelligent scaffolding support novices across the science ecosystem?}

% -- The current study mainly explored the review writing process, there are other stages of conference peer review that novices found difficult: eg. writing rebuttal. Scaffolding can be used in different peer review stages: writing review, meta-review, rebuttal
Our study of ReviewFlow indicates novice reviewers can benefit from intelligent scaffolding: it helps them evaluate the submission and write more comprehensive reviews. Participants mentioned how they gradually learned about the research community's practices and became more confident in their review writing. Beyond just writing reviews, the strategies for intelligent scaffolding built into ReviewFlow have the potential to provide value to other aspects of the peer review ecosystem~\cite{shah2018design}. For example, similar intelligent scaffolding can be used to support novice reviewers to revise their papers and write a persuasive rebuttal. The process of meta-reviewing, or summarizing independent reviews, could also benefit from scaffolding since there are new people jumping into that role each cycle~\cite{metawriter}.

% concerns of deployment of the system in the peer review community, think of pushback and concerns from the general community.  
% add point about future deployments within a research community... what commitment it would take, how robust would the system need to be, what validation would it provide, how would it be maintained... maybe you don't need to answer these, but rather say these are key considerations.
Future deployments of this type of intelligent scaffolding would require careful consideration and round-table discussions within the research community. Previous research revealed that writing with an opinionated language model can affect participants' attitudes towards social topics in writing~\cite{jakesch2023co}. In our study, even though our scaffolding was carefully designed to allow the user to drive the process and to avoid biasing the decision, we still heard participants who were concerned about other people using such a system. Interestingly, very few participants were concerned that their own decision or writing was being biased by the AI, instead, they worried about how others would appropriate the intelligent support and how it might erode a community's trust in the peer review system. To deploy the system in the future and build trust with people in the community, we need to make sure that the system is robust and trustworthy~\cite{wang2023decodingtrust}. A real-world deployment would need to have a large-scale consent process and commitment from key stakeholders. 

% -- Scaffold research paper reading, searching, sensemaking and writing(AI2’s work)
% (move the first paragraph below... expand on the second to include pointers to work by AllenAI, and others, to support lit reviews, to support critical reading, to support idea generation across scientific displines, etc.)
Beyond its application in peer reviewing, intelligent scaffolding could support a range of of complex knowledge work in science, such as paper reading, literature reviewing, sense-making, and paper writing~\cite{palani2023relatedly,liu2023selenite,zhang2023visar,lin2024rambler}. Notable projects like the semantic reader project have developed interactive and dynamic reading interfaces to aid paper reading and citation discovery~\cite{lo2023semantic,kang2022threddy,august2022paper,fok2023scim}. These studies suggest broader possibilities for incorporating intelligent, process-driven scaffolding into science work.

\subsection{Ethical considerations of AI scaffolding for academic review}
%  be more explicit about the potential concern and each mitigation strategy. You list one:  auto-generating bad reviews can be mitigated by limiting the stages of the process supported by intelligent scaffolding.  
% But what else?  What about concerns that other people might be using LLMs to write reviews? Any mitigation for that?  What about misinformation or other mistakes that might artificially sway people? For that one, we could talk about building in "reflective skepticism" into any tools that leverage LLMs. 

Generative AI and LLMs introduce numerous ethical considerations in the design of human-AI collaboration systems. In the context of academic publishing and peer review, a primary concern involves the potential violation of academic integrity and harm to authorship when directly incorporating automatically generated content into writing artifacts. 
We try to mitigate this in ReviewFlow by constraining the use of LLMs to primarily sensemaking activities, like guided note-taking. In the final step, ReviewFlow synthesizes an outline, not paragraphs, even though LLMs are quite capable of doing so. To limit data sharing, ReviewFlow only sends the LLM the reviewer's notes and highlighted segments, not the full paper. While individual reviewers may still choose to (mis)use LLMs in these ways, ReviewFlow subtly prioritizes learning and thinking over getting the job done fast. 

Another concern revolves around  the tendency of LLMs to create inaccurate information or to mislead people\cite{huang2019reducing,jakesch2023co}. In line with our scaffolding strategy, ReviewFlow includes a checklist pop-up that allows users to engage in self-reflection, with reminders to proofread and fact-check their review. To enhance transparency in the writing process, future systems might explore the development of intelligent highlighting. For example, highlights could directly color-code the portions that were user-generated and those that were automatically generated, providing a clear visual indication and promoting a more transparent and accountable writing process.

Given the uncertainties, future LLM-powered tool designers should consider strategies for inducing users into a ``reflective skepticism'' around all data, especially those produced by machine models. As exemplified in prior work ~\cite{lai2011critical}, 
this involves fostering a mindset of critical evaluation and thoughtful questioning to counteract the potential limitations, hallucinations, biases, or misinformation that may arise from these language models.

\section{Limitations}
% 7.4. Limitation of this study
% -- Only use short papers in HCI
% -- Have a small sample of participants
% -- LLM has limited ability, we are using GPT4, with the development of LLM, it can improve the experience in general, but more challenges also need to be addressed. Eg. hallucinations may be not easy to capture, and need fact-checking

Our study has several limitations. First, ReviewFlow combined multiple scaffolding strategies into one tool, leaving future work to understand to what extent each strategy impacted the outcomes. Ablation studies could provide more insight into the effectiveness of each scaffold.  For instance, a study with three experimental conditions, each providing only one type of scaffolding would allow for a more comprehensive evaluation.
Second, we simulated a mock peer review scenario where users had about one hour to read and write a review for a short paper. In practice, as we learned in our preliminary interviews, writing peer reviews takes hours. A longer study could give insights into the enduring value of scaffolding on a longer paper.
Third, we selected two papers in HCI to use in our experiment, but different research domains and communities have different guidelines and standards for reviewing. Deploying the system across research communities may bring new insights into the generalization of scaffolding strategies. Lastly, the underlying machine models and LLMs will keep improving, which can impact the performance of ReviewFlow, for better or worse. Continuous updates and adaptations to the latest AI models would be required to maintain the tool's effectiveness and relevance.

\section{Conclusion}
This research explores techniques for integrating LLMs into intelligent scaffolding for academic peer reviews. Our formative studies found that expert reviewers adopted a workflow of annotating, note-taking, and synthesizing notes before writing, while novices lacked perspective on the prior work in the domain and reviewing standards. Modeling the expert workflow, we developed ReviewFlow --LLM-supported workflow that scaffolds novices using contextual reflection cues, in-situ knowledge support, and notes-to-outline synthesis. In a within-subject experiment with 16 novice reviewers, we found that ReviewFlow led to more comprehensive peer review and higher self-efficacy on the task. We further discuss the implication of intelligent scaffolding in knowledge work.

\bibliographystyle{ACM-Reference-Format}
\bibliography{ref}

\pagebreak
\section*{Appendix}

\begin{table*}[htb]
    \centering
    \begin{tabular}{|p{0.12\textwidth} | p{0.45\textwidth} | p{0.4\textwidth} |}\hline
         \textbf{Features} &  \textbf{Prompt} & \textbf{Examples}\\ \hline
         Section-level contextual cues  & \textbf{System role:} You are an expert peer reviewer providing guided questions for novice peer reviewers to help them write better reviews.  & \textbf{Importance:} Is the problem addressed in the submission, concerning the video explicitness of a video conferencing interface for virtual\\
         & \textbf{Context}: Given this paper’s abstract: \{ abstract \} & studying, important to the CSCW community? \\
         & \textbf{Template}: What could be a helpful guided question for novice reviewers to evaluate the \{ aspect \} of this paper for this section of the paper? Please output 1 concise question with a maximum of 25 words &\textbf{Novelty:} How does this study relate to existing work in the field of virtual social activities and video streaming? Does it introduce a novel approach?\\
         phrase-level contextual cues  & \textbf{System role:} You are an expert peer reviewer providing guided questions for novice peer reviewers to help them write better reviews.  & \textbf{Validity:} Does the Method section provide clear and well-justified explanations of the research prototype's design, including the choice of video \\
         & \textbf{Context}: Given this paper’s abstract: \{ abstract \} & versions and activity recognition method?\\
         & \textbf{Template}: What could be a helpful guided question for novice reviewers to evaluate the \{ aspect \}  of this paper for this specific paragraph:\{ paragraph \}. Please output 1 concise question with a maximum of 25 words for this specific paragraph. & \textbf{Clarity:} How comprehensive does the method explain the recruitment process, demographics, and procedures of the user study, ensuring clarity for a CSCW audience? \\ \hline
         Notes-to-outline synthesis & \textbf{System role:} I am building a web application that helps novice peer reviewers write better peer reviewers. Your task is to synthesize the reviewer's notes into an outline. Please provide an outline with sections of the summary of the paper, strengths, and weaknesses. To accomplish this task, I am providing an abstract of the reviewed paper, each of reviewer's notes and corresponding contents of the reviewed paper for context, and the topics that should be used to generate the weakness part of the outline. & \textbf{Summary:} \hfill \break \tabitem Explored new design strategies for virtual studying focusing on sharing presence. \hfill \break \tabitem Context of the study is the advent of COVID-19 and arising of new virtual social activities. \hfill \break \tabitem Investigated minimizing video explicitness in virtual studying using various strategies. \hfill \break \tabitem The study involved four virtual studying teams that used prototypes and participated in interviews.\\
         & \textbf{Template: }Here is the paper abstract:\{abstract\}; Here are the user annotations that contain Note and paper content only as context:\{notes\}. Please create three to five concise bullet points for each section (strengths, weaknesses) out of the reviewer's notes for the templated outline. The weakness part of the outline should be based on the topics for weakness but with detailed descriptions using only the notes under each topic. The topics should have a length of at most 10 words. The details should have a least 10 words and be a complete sentence. &
\textbf {Strength:} 
\hfill \break \tabitem Relevance to current societal changes due to COVID-19. 
\hfill \break -- Paper timely addresses the shift to virtual activities sparked by COVID-19. 
\hfill \break \tabitem  Addresses an emerging need in virtual collaborative environments.  
\hfill \break -- Investigates novel video conferencing interfaces for enhanced privacy in virtual studying. 
\hfill \break \textbf{Weakness: }
\hfill \break \tabitem   Limited scope of the study sample.
\hfill \break -- Sample size is limited to four virtual studying teams, restricting broader applicability of findings. 
\hfill \break \tabitem  Lack of precision in relating strategies to prior research. 
\hfill \break -- The paper lacks precise connections between its strategies and the existing research, undermining its innovative claim. 
\hfill \break \tabitem  Insufficient citations on related work, specifically Barbara's contributions 
\hfill \break -- Needs more references to Barbara's work to establish the paper's context. 
\\ \hline
    \end{tabular}
    \caption{Prompts and examples used in the ReviewFlow system.}
    \vspace{-2em}
\end{table*}

\end{document}